\documentclass[doublespace]{WileyMSP-template}
\usepackage{multirow}
\usepackage{graphicx}
\usepackage{longtable}
\usepackage{lipsum}
\usepackage[table,xcdraw]{xcolor}
\usepackage{rotating}
\usepackage{lineno}
\usepackage{ragged2e}
\begin{document}
\pagestyle{fancy}
\title{Topological Insulators: An in-depth Review of their Use in Modelocked Fiber Lasers}
\maketitle


\author{Shyamal Mondal*},
\author{Rounak Ganguly} and
\author{Krishnakanta Mondal}


\dedication{}

\begin{affiliations}
Dr. Shyamal Mondal\\
Department of Electronics and Communication Engineering, SRM Institute of Science and Technology, Tamil Nadu, 603203, India\\
Email Address: shyamal.kgec@gmail.com \\

Mr. Rounak Ganguly\\
Department of Electronics and Communication Engineering, SRM Institute of Science and Technology, Tamil Nadu, 603203, India\\
Email Address: ganguly.rounak609@gmail.com

Dr. Krishnakanta Mondal\\
Department of Physics, Central University of Punjab, Bathinda, 151001, India,\\
Email Address: krishnakanta1987@gmail.com \\

\end{affiliations}



\begin{abstract}
\justify
Topological Insulators (TIs) exhibit exciting optical properties, which opens up a new pathway to generate ultrashort pulses from fiber lasers. Layered TIs display distinct saturable absorption property due to excited state absorption, as compared to their bulk structures.  Moreover, the electronic structures of the films of TIs depend on the thickness of the films due to the quantum confinement of the electrons. By virtue of this, the nanoparticles of TIs play a key role in all-fiber modelocked laser. By tweaking the crystal structures of TIs, it is possible to generate ultrashort pulses across the visible, near-infrared and mid-infrared wavelengths. Starting from the crystal structures and density of states calculations, how different topological insulators can be fabricated and integrated as an efficient passive saturable absorber in all-fiber modelocked lasers with the capability of producing fundamental to high-harmonic pulse generation are described clearly in this review report. Moreover, this report reviews the current state-of-art of TI-based saturable absorbers and their applications in different regimes of modelocked fiber lasers.

\end{abstract}

\tableofcontents
\section{Introduction}

\justify
Ultrafast science is a vibrating and emerging multidisciplinary field of research today. Femtosecond fibre lasers are the central building blocks of numerous photonic systems used in communication, defense, medical and industrial applications.
All-fiber ultrafast femtosecond lasers are favorable over free-space solid-state lasers as they are robust, portable, cost-effective and can be coupled with other optical fiber systems to deliver ultrashort pulses at remote places. 
Before 21$^{st}$ century, people were mostly dependent on passively modelocked solid-state lasers because of their ability to deliver high peak power and shorter pulsewidth. 
However, with the availability of high-power pump laser diodes at the gain fibers’ absorption lines and due to the technological advancements of new 2D nanomaterials, the laser community become incline towards ultrafast fiber laser. 
These 2D-materials are used as saturable absorbers (SA) for their unique optical properties to create the ultrashort  pulses.

Saturable absorber (also known as modelocker) is the key component of ultrashort pulse lasers \cite{ haus1975theory}. They are used for in-phase locking of the longitudinal modes of a laser cavity. The ultrashort pulsewidth mainly depends on the electronic and optical properties of SA. 
Due to the Dirac like electronic structure of graphene which is discovered in 2004  \cite{novoselov2004electric}, emerges as a candidate for broadband saturable absorption. 
However, graphene has small absorption rate and low value of modulation depth \cite{chang2010multilayered} which limits it to be an excellent saturable absorber.
This motivates the researchers to explore how other Dirac materials would work as a SA having better absorption properties.
As a consequence, several classes of 2D and nearly 2D materials like, Transition Metal Dichalcogenides (TMDs) \cite{ bikorimana2016nonlinear}, Black Phosphorous (BP)  \cite{sotor2015black}, Topological Insulators (TIs)  \cite{bernard2012towards}, etc. have emerged to meet the expectation.
In particular, recently developed TIs as saturable absorber have drawn more attention due to their rare electronic structures and optical properties. 
Therefore, in this report, our discussion is solely focused on TI and its application in ultrafast fiber laser. Moreover, we have only considered few-layered TIs (nearly 3D TIs) as saturable absorbers for our review article.

\subsection{Topological insulators}
Topological insulators are insulating materials having spin-orbit coupling so strong that the ordering of the bands about the insulating bulk gap are inversed. 
The bulk properties of these materials are not so different from other insulators, but the presence of exotic excitation states on their surfaces earn them the ‘topological’ classification  \cite{beindenkopf2013visualizing}.
The presence of strong spin-orbit coupling-induced band inversion and time-reversal symmetry make the surface of TIs conducting and the interiors insulating. 
Such interesting feature cannot be seen in ordinary insulators. 
The TI has a better metallic boundary arising from topological invariant, compared to any other insulator.
It has been shown that edge states occur at these boundaries in which spin up and spin down electrons are in the integer quantum hall effect like states, each feeling opposite effective magnetic fields arising from spin orbit coupling \cite{hasan2010colloquium}.
It has been theoretically predicted and experimentally demonstrated that the surface states have electrical properties that are fundamentally different from other 2D conducting states discovered so far.
In 2005, Kane and Mele \cite{kane2005z}, through their theoretical studies, discovered a new topological invariant that could mathematically be calculated for any TI and would help predict if it has a stable edge state.
It allowed them to show that stable edge states are possible even without the presence of magnetic field; the resultant 2D state was the first TI to be understood. 3D TIs were first discovered by Hsieh et al.  \cite{hsieh2008topological} in 2008 where they studied crystal structures of bulk bismuth-antimony alloy.
Subsequently, Bismuth telluride (\uppercase{B}i$_2$\uppercase{T}e$_3$), bismuth selenide (\uppercase{B}i$_2$\uppercase{S}e$_3$), antimony telluride (\uppercase{S}b$_2$\uppercase{T}e$_3$), etc. were all discovered \cite{zhang2009topological, li2011electronic, betancourt2016complex,lawal2017sb2te3,jhon2018topological}.

\begin{figure}[h!]
\centering\includegraphics[width=12cm]{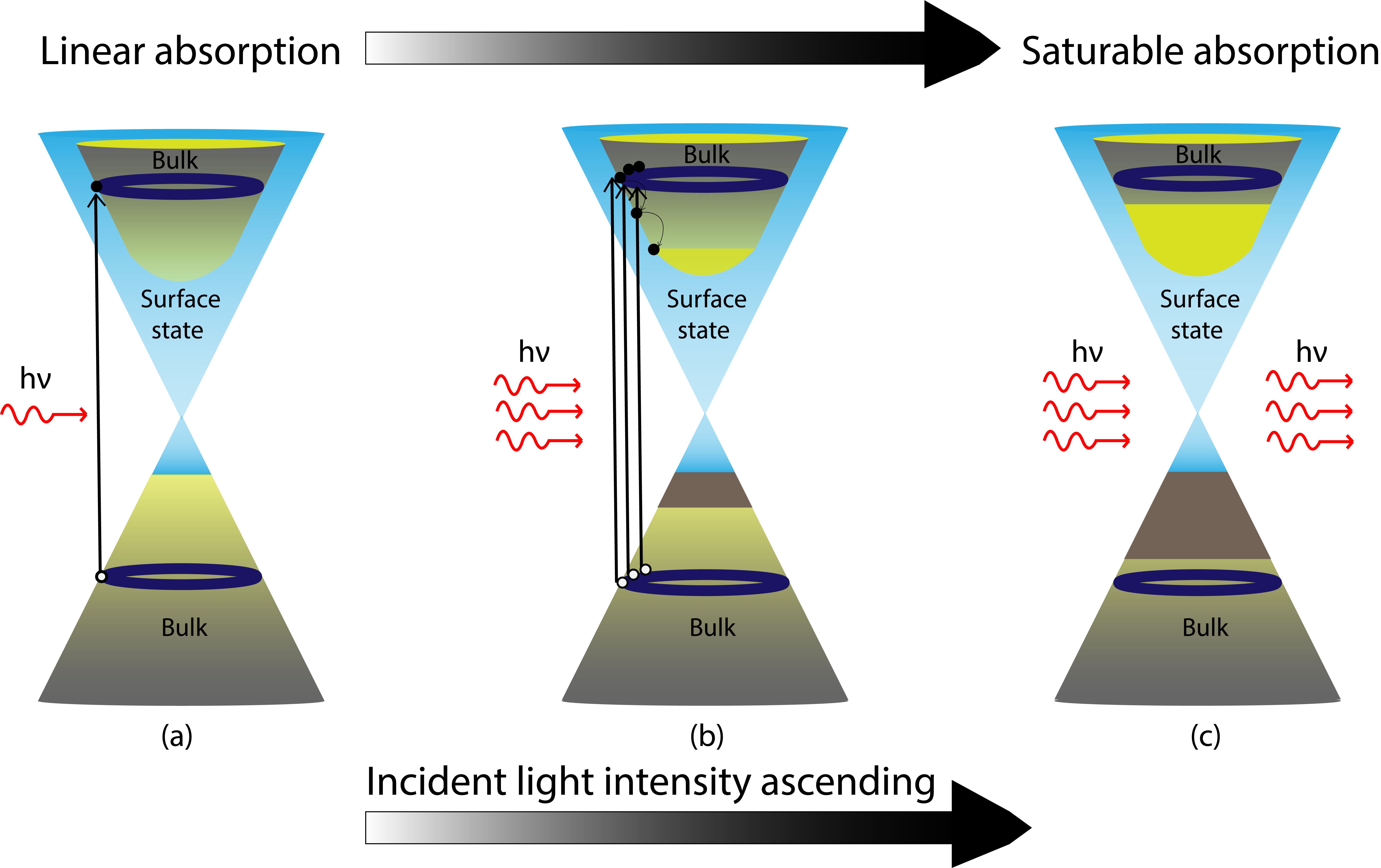}
\caption{Mechanism of saturable absorption in TISA. (a) Linear absorption; (b) Bleached condition; (c) Saturable absorption and light transmission} 
\label{elecstruc}
\end{figure}

The saturable absorption of topological insulators can be explained through Fig.\ref{elecstruc}. In TI, both conducting surface states and insulating bulk states are present. During the start of a laser, when intracavity intensity crosses the threshold, the saturable  absorber starts its job. The noise-like pulses having low intensity falling on the SA excite all the corresponding electrons in the valence band of surface and bulk states to the conduction band of bulk states, making the ground states empty (Fig.\ref{elecstruc}(b)). At this stage, TISA cannot absorb further intracavity photons and becomes transparent, allowing all the incoming photons to simply pass through (Fig.\ref{elecstruc}(c)), which is called saturable absorption. Because of this phenomena, the tip of the strongest noise eventually escapes from the SA and acquires gain over the roundtrips due to the stimulated emission process. Therefore, the wings of the strongest noise always get absorbed by the SA, thus making it narrower. A high intensity pulse is formed by constructive interference between the in-phase longitudinal cavity modes. As many modes are in-phase, the resulting pulse becomes shorter, peak power gets higher and the corresponding frequency spectrum becomes broader. Here TISA acts as a gate which opens whenever the high intensity pulses reach it and blocks any low intensity (noise-like) pulses.

Topological insulators have surface band structures similar to graphene, so they were extensively researched  \cite{hsieh2008topological}. Bernard and Zhang  \cite{bernard2012towards}, studied the nonlinear optical response of TIs and found out that they show saturable absorption at telecommunication wavelengths. 
Several studies have reported that TIs have better modulation depths and more absorption rates than graphene for specific wavelengths  \cite{martinez2013nanotube, yan2015practical}.
The detailed study of structural and electronic properties of TIs is given in the preceding section.

\subsection{Topological insulator saturable absorber based fiber lasers}
\subsubsection{TISA in Erbium-doped fiber laser}
\begin{figure}[h!]
\centering\includegraphics[width=14cm]{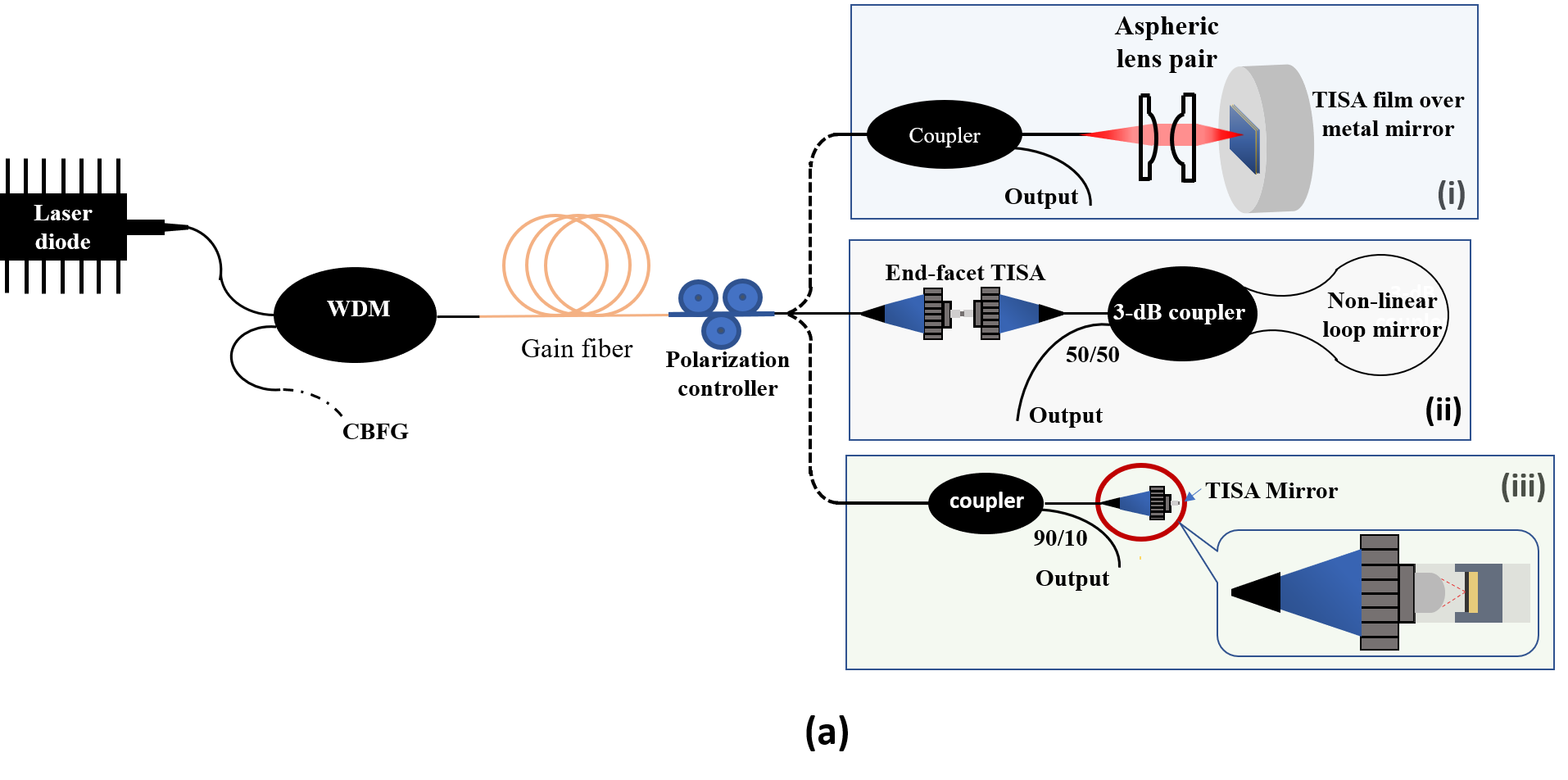}
\centering\includegraphics[width=14cm]{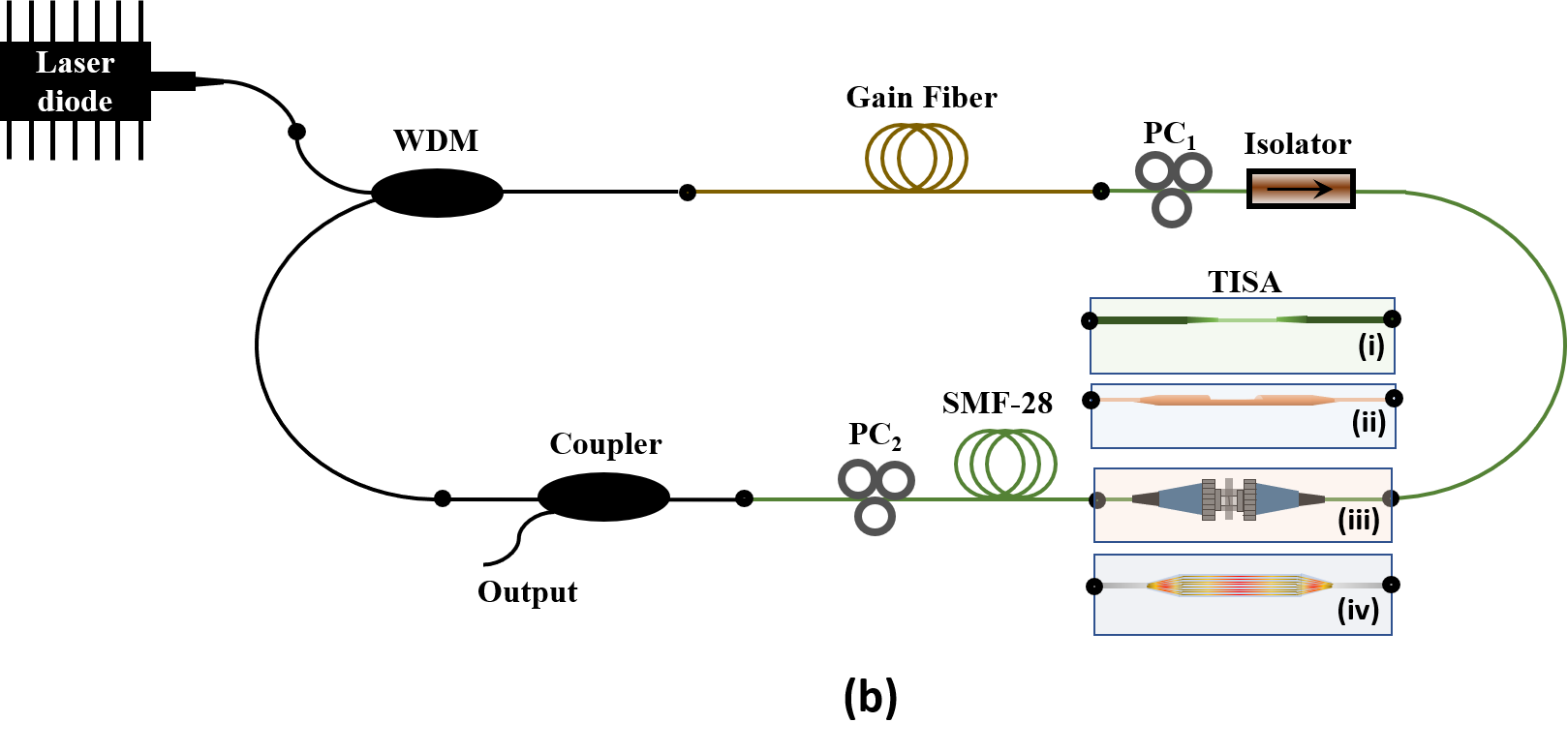}
\caption{Schematic diagram of modelocked fiber lasers in (a) linear cavity configuration with (i) TISA film over metal mirror, (ii) end-facet TISA and (iii) TISA mirror at end-facet of the fiber; (b) ring cavity configuration with (i) tapered fiber TISA, (ii) side-polished TISA, (iii) end-facet connectorized TISA and (iv) PCF-TISA.}
\label{FL_linear}
\end{figure}

Active research work on topological insulator based saturable absorber (TISA) and its application in fiber lasers have been in the run since 2012. 
TISA can be used in both linear as well as in ring cavity fiber laser configuration. Fig. \ref{FL_linear} shows the schematic diagram of the same, where TISAs are used at different positions in fiber laser resonators.
\begin{itemize}
 \item \textit{{Review of fundamental mode-locking}} 
\end{itemize}    
    
Optical properties exhibited by TIs were first investigated by C. Zhao et al. \cite{zhao2012ultra} in early 2012 in the form of a modelocker devised for ultrafast laser generation. 
They successfully developed a 1.55 $\mu$m modelocked erbium-doped fiber laser (EDFL) based on Bi$_2$Te$_3$ SA with 1.21 ps pulses. 
Their investigation opened the doors for TIs in the field of photonics, especially in broadband saturable absorption applications. 
TIs straightaway showed better performance than graphene based saturable absorbers.
Simultaneously, the same core group worked with Bi$_2$Se$_3$ \cite{zhao2012wavelength} to test its saturable absorption capabilities at $\sim$1560 nm. 
To this list of TI based SA, another material named Sb$_2$Te$_3$ was added by J. Sotor and his co-workers  \cite{sotor2014mode}.
This group achieved a remarkable feat by generating a TISA based stretched-pulse modelocked fiber laser for the first time, with an exceptional pulse width of $\sim$128 fs and an average power of 1 mW \cite{sotor2014sub}. 
In the year 2014, another group of researchers found bulk p-type Bi$_2$Te$_3$ in a thermoelectric cooler and polished it to get nanosheets, which was incorporated at end-facet fiber to have modelocked EDFL \cite{lin2014soliton}. 
They achieved pulse width of minimum $\sim$403 fs havng spectral bandwidth 6.86 nm, with a repetition rate of 28.5 MHz. However, the information regarding modulation depth and output power are doubtful. Moreover, the data of threshold pump power and Signal-to-noise ratio(SNR) are not available. 
J Boguslawski et al. \cite{boguslawski2014mode} realized an EDFL using Sb$_2$Te$_3$ SA developed by liquid phase exfoliation (LPE) at $\sim$1550 nm which successfully achieved a pulse duration of $\sim$450 fs.
It was shown by H. Liu et al. \cite{liu2014femtosecond} that polyvinyl alcohol could be an excellent host material for fabricating high performance TISAs.
Moreover, they achieved a pulse width of $\sim$660 fs which demarcates the experiment as a success. 
Few other noteworthy published works in 2014 were also found in the same domain \cite{chen2014formation,luo2014observation}.

In 2015, J. Boguslawski et al.  \cite{boguslawski2015dissipative} set a record by generating dissipative soliton of EDFL operating in near-zero dispersion regime with a TISA where they generated sub-170 fs pulses at $\sim$1.55 $\mu$ m. 
The following year, a novel work was demonstrated by Hou-Ren Chen \cite{chen2016high}, where they developed a reflection type Bi$_2$Te$_3$-gold film-based nonlinear optical modulator fabricated using pulsed laser deposition (PLD).
The pulses were in the femtosecond order ($\sim$452 fs) and showed stable modelocked operation for over 24 hours.
The shortest pulse duration with TISA till date has been achieved by W. Liu et al. \cite{liu201670} at $\sim$1.55 $\mu$ m with Sb$_2$Te$_3$. They reported 70 fs pulse width with a spectral bandwidth of 63 nm and also having a maximum output power of 63 mW.
The following year, Yanhua. Xu et al. \cite{xu2017bilayer} developed high quality bi-layer Bi$_2$Se$_3$ nano-platelets based SA using simple and efficient bottom-up synthesis method and achieved 579 fs pulses.
B. Guo et al. \cite{guo2018sub} once again tested a TISA at 1.55 $\mu$m region. However, this time they demonstrated a new topological insulator called bismuthene, a hexagonal crystal 2D layer of bulk Bismuth crystal, which generated pulse duration as short as 193 fs using EDFL. 
In 2019, Qin Wei et al. \cite{wei2019large} achieved a milestone by fabricating Bi$_2$Te$_3$ SA using chemical vapour deposition (CVD) method for the first time for saturable absorption at $\sim$1.55 $\mu$m wavelength radiation and a maximum average output power of 40.37 mW with 23.9 nJ pulse energy. 
One of the lastest works on TISA was published by G. Liu et al. \cite{liu2020ultrathin}, where they demonstrated another new TI named tellurene, a 2D layer of tellurium atoms arranged in hexagonal array, fabricated using LPE. 
A pulse duration of $\sim$1 ps and average output power of 3.69 mW  when pumped with 314 mW, highlighted the success of tellurene in ultrafast photonics applications. 
Moreover, another new TI material, AuTe$_2$Se$_{4/3}$ was introduced in this field by W. Liu et al. \cite{liu2020ultrafast}. 
This material was fabricated using self-flux method and its SA gave promising results when placed in a modelocked EDFL. 
Pulses as short as 147.7 fs and spectral bandwidth of 30.26 nm definitely claimed its saturable absorption capabilities and made it a viable SA. 
So far, the literature reviews were concentrated on fundamental modelocking of fiber lasers at different wavelengths. Researchers have also achieved dual wavelength modelocked pulses simultaneously from the same fiber laser cavity. 

\begin{itemize}
 \item \textit{{Review of multi-wavelength mode-locking}} 
\end{itemize}

Till now only fundamental modelocking with TISA based fiber laser at $\sim$1.55 $\mu$m has been discussed, but we also see multiple-wavelength generation using TISA.
By controlling the fiber birefringence using polarization controller, the intra-cavity intensity can be distributed among the different polarization states, which helps to achieve multi-wavelength modelocking regime. 
The origin of the dual and multi-wavelength modelocking as well as harmonic modelocking have been described in the section 5. 
Using this principle, B. Guo et al. achieved two orthogonally polarized fundamental modelocked pulses at $\sim$1565.5 nm and $\sim$1566.3 nm wavelengths respectively at pump power 97.3 mW  \cite{guo2015observation}. 
They also realized a multi-wavelength ultrafast EDFL with the aid of a Bi$_2$Se$_3$-PVA film-based SA  \cite{guo2015topological}. 
Moreover, few months later the same group published a work on dual-wavelength pulse in a TISA based EDFL cavity centered at 1561.60 nm and 1562.10 nm \cite{guo2015dual}.

Later, another group has also demonstrated dual-wavelength soliton generation at 1532 nm and 1557 nm using a Bi$_2$Se$_3$-PVA film-based SA \cite{li2017analysis}.

\begin{itemize}
 \item \textit{{Review of harmonic mode-locking}} 
\end{itemize}
To achieve higher repetition rate, harmonic modelocking (HML) has also been realised using TISA based modelocked fiber laser. 
First ever attempt to generate higher repetition rate by utilizing harmonic modelocking states was shown by Luo et al. \cite{luo20132} where they achieved 2.04 GHz resulting from 418$^{th}$ harmonic of fundamental cavity frequency (4.88 MHz). It successfully demonstrated higher nonlinearity and saturable absorption capabilities at $\sim$1.55 $\mu$m of a micro-fiber based TISA. The very next year, J. Sotor et al. \cite{sotor2014harmonically} also demonstrated harmonic modelocking states with mechanically exfoliated Sb$_2$Te$_3$ based SA integrated at the fiber-facet. They achieved 304 MHz repetition rate by producing 81$^{st}$ harmonic by adjusting the birefringence and pump power. 
M. Liu et al. \cite{liu2014dual} went one step further to produce dual-wavelengths along with harmonic modelocking by using Bi$_2$Te$_3$ TISA. 
Repetition rates of 388 MHz and 239 MHz were generated at 1557.4 nm and 1559.4 nm respectively from EDFL.
Researchers from Shenzhen Key Laboratory of Laser Engineering \cite{yan20152} developed microfiber based TISA by incorporating Bi$_2$Te$_3$ thin-films in tapered fiber, which exploits the HML characteristics (upto 2.95 GHz) due to excess nonlinearity. Ultrashort pulses of $\sim$320 fs was successfully generated at $\sim$1.55 $\mu$m along with a maximum output power of 45.3 mW.
J. Lee et al. \cite{lee2015femtosecond} had also worked on harmonic modelocking with an EDFL ring cavity and had generated 773.85 MHz.
 
At the end of 2015, L. Gao et al. \cite{gao2015stable} accomplished two different experiments using Bi$_2$Se$_3$ based SA at $\sim$1.55 $\mu$m range, to demonstrate its saturable absorption capability. One experiment was conventional soliton modelocking using HML technique and the other was dissipative soliton modelocking in a normal dispersion cavity. 
Next year, Jian Ping Li et al. \cite{li2016vector} reported the highest ever frequency achieved with the aid of harmonic modelocking. 
They used Bi$_2$Se$_3$ with tapered fiber in an EDFL to demonstrate up to 356$^{th}$ order of HML which generated 3.57 GHz with femtosecond pulses having width of 824 fs.

Very recently, the second highest repetition rate of 3.125 GHz with 200$^{th}$ harmonic order has been achieved by HML technique at $\sim$1.55 $\mu$m \cite{jin20183}.
Another similar work was demonstrated by Z. Wang et al. \cite{wang2019generation} who reported HML of 46$^{th}$ order at 607.2 MHz at $\sim$1.563 $\mu$m using Sb$_2$Te$_3$ on microfiber SA.

Apart from C-band wavelength range, TISA has also been used to modelock the fiber laser at long wavelength regime (L-band). 
The first TISA in L-band was reported by 
K. Li et al. \cite{li2015band} adopting Bi$_2$Se$_3$-PVA film-based SA for EDFL. They successfully generated pulses as short as 360 fs centered at $\sim$1.6 $\mu$ m, which is the shortest pulse generated till date in the L-band regime.
In the same year, Y. Meng and his group \cite{meng2015high} reported a Bi$_2$Se$_3$ based tapered fiber TISA incorporated in a Er:Yb co-doped double clad fiber laser for L-band regime ($\sim$1.6 $\mu$m). 
They achieved 640.9 MHz repetition rate (91$^{st}$ harmonic order) and maximum output power of 308 mW, which is reported as the highest output power generated using TISA $\sim$1.6 $\mu$m  till date. 
TISA has also been used in yetterbium and thulium-holmium doped fiber lasers to generate pulses in 1 $\mu$m and 2 $\mu$m wavelength range respectively.

\subsubsection{TISA in Ytterbium-doped fiber laser}
Ytterbium-doped silica fibers are commonly used as gain mediums to generate radiation in near infrared regime (typically, 1030 - 1080 nm). 
Yb-doped modelocked fiber lasers (YDML) have been currently applied in the industrial market heavily. 
These lasers provide a large number of advantages such as high average powers and efficiencies, small footprint and superior beam quality. 
Compared to Nd:YAG lasers in this wavelength regime, apart from short pulse duration, YDML provide better pulse controlling and shaping, low maintenance work, higher repetition rate which greatly enhance the industrial production \cite{zervas2014high}.
In the year 2014, for the first time, Bi$_2$Se$_3$ based side-polished fiber in YDML setup was shown by C Chi et al. \cite{chi2014all}. Here, they used $\sim$17 $\mu$m-thick Bi$_2$Te$_3$ TI film and achieved $\sim$230 ps pulses. 
In the same year, Z. Dou et al. \cite{dou2014mode} achieved pulse width of 46 ps and the repetition rate of 44.6 MHz using Bi$_2$Se$_3$ based TISA at 1030 nm.
After this report, in the following year, another group has demonstrated modelocking with Bi$_2$Se$_3$ TISA at 1052 nm in all-normal-dispersion regime  \cite{lu2015yb} and achieved repetition rate of 9.8 MHz, 3-dB spectral width of 1.245 nm and pulse duration of 317 ps. 
The latest report of TISA in YDML came in  2017 by Lu Li et al. \cite{li2017high} where they achieved a repetition rate of 6.2 MHz with Bi$_2$Te$_3$ TI at 1063.4 nm

\subsubsection{TISA in Thulium/Holmium-doped micron fiber laser}
Currently, 2 $\mu$m fiber laser is drawing attention due to its wide applications in medical surgeries, gas sensors, material processing, remote sensing, molecular spectroscopy  \cite{scholle20102}, etc. 
Upon doping of silica fiber with thulium and/or holmium rare earth elements, while pumping with suitable wavelength, it emits radiation in 2 $\mu$m regime. 
The first thulium doped fiber laser (TDFL) was discovered by Hanna et al., in 1988 using a 797 nm dye laser as the pump source and the first 2 $\mu$m Q-switched TDFL was developed by Myslinski et al. \cite{myslinski1993self} in 1993 by using an acousto-optic modulator. 
To modelock at 2 $\mu$m regime, it is favourable that the saturable absorber contains dense energy levels having bandgap energy of 0.6-0.65 eV. Using single-layered TMD materials, such as MoS$_2$, WS$_2$, WSe$_2$, etc. it cannot achieve the required bandgap energy directly for modelocking at 2 $\mu$m regime. 
On the other hand, topological insulators due to its inherent structural property, is suitable for Th-Ho-doped modelocked fiber laser and can directly generate 2 $\mu$m wavelength pulses.

In 2014, M. Jung et al. \cite{jung2014femtosecond} used $\sim$30 $\mu$m-thick sheets of Bi$_2$Te$_3$ TISA to achieved $\sim$795 fs pulses at 1935 nm. 
A year later, Ke. Yin et al. \cite{yin2015soliton} reported existence of both bunched solitons as well as harmonically modelocked solitons in Bi$_2$Te$_3$ SA based thulium/holmium co-doped fiber laser at $\sim$2 $\mu$m for the first time. 
In bunched soliton modelocking, they achieved maximum 15 pulses in a single bunch at fixed position of the polarization controller and increasing the intra-cavity gain.  
After that, another group investigated Bi$_2$Se$_3$ based SA in Tm-doped fiber laser at 2 $\mu$m region  \cite{lee2018femtosecond}.
These researchers successfully generated $\sim$853 fs pulses at 1912 nm wavelength, which established the compatibility of TISA in the eye-safe zone of laser radiation. 

Therefore, it has seen that TISA is an excellent candidate as a modelocker in comparison to other 2D-nanomaterial or few layers' based SAs such as graphene, various transition metal dichalcogenides (MoS$_2$, WS$_2$, MoSe$_2$, WSe$_2$, ReSe$_2$, etc.), black phosphorous, etc.  \cite{liu2017emerging,wu2018high, ma2019recent}. The major advantage of TISA is that it can be used to modelock almost all available fiber laser lines at near-infrared to mid-infrared wavelength regimes.
Other advantages are ease of fabrication and integration to the fiber laser cavity, more stable, higher damage threshold, cost-effective, less prone to foreign materials, can modelock in dissipative soliton, bound soliton, bunched soliton, all-normal dispersion and few other regimes. 
From the close study of these works, the importance of TI as an excellent choice of modelocker is clearly understood.

Among the real SAs only semiconductor saturable absorber mirror (SESAM) has been commercialised in the market. However, either they are bulky or mounted at the tip of a fiber connector, only working in reflected mode. Therefore, it cannot be paired with an all-fiber mode-locked ring laser. On the contrary, the advantage of a TISA is that it can be integrated in a fiber laser in several ways (such as, side-polished, tapered, between the connectors and inside micro-structured fiber). 
TISA can also be tuned over different wavelength ranges unlike a SESAM, which has a  fixed working wavelength. In addition, multi-wavelength and harmonic mode-locking regimes are possible to achieve by TISA. From an economical perspective too, a TISA is much cheaper to fabricate than buying a SESAM.

The structural and electronic properties of TIs to exploit the best modelocking scenario has been described in the next section. 
The fabrication-integration techniques and the characterizations of TISAs have been depicted in section 3 and 4, respectively.
A brief discussion of the origin of different types of modelocking using TISAs have been illustrated in section 5. The 'Challenges and future directions' of this \textit{'TISA-based modelocked fiber laser'} research has been outlined in section 6.
Finally, conclusion with futuristic pathway in this field has been specified in the last section.

\section{Structural and electronic properties of topological insulators}
The well-known topological insulators with formula Y$_2$X$_3$ (Y = Bi, Sb, X = Te, Se) belong to the trigonal lattice with R$\bar{3}$m rhombohedral crystal symmetry.
The bulk structure of Y$_2$X$_3$ crystal comprises of three quintuple layers along c-axis (0001 plane). These layers are connected to each other by feeble Van der Waals force. Each quintuple layer consists of five crystal layers with the pattern X1—Y—X2—Y—X1. 
The individual crystal layer shows hexagonal symmetry like graphene. 
\begin{figure}[h!]
\centering\includegraphics[width=14cm]{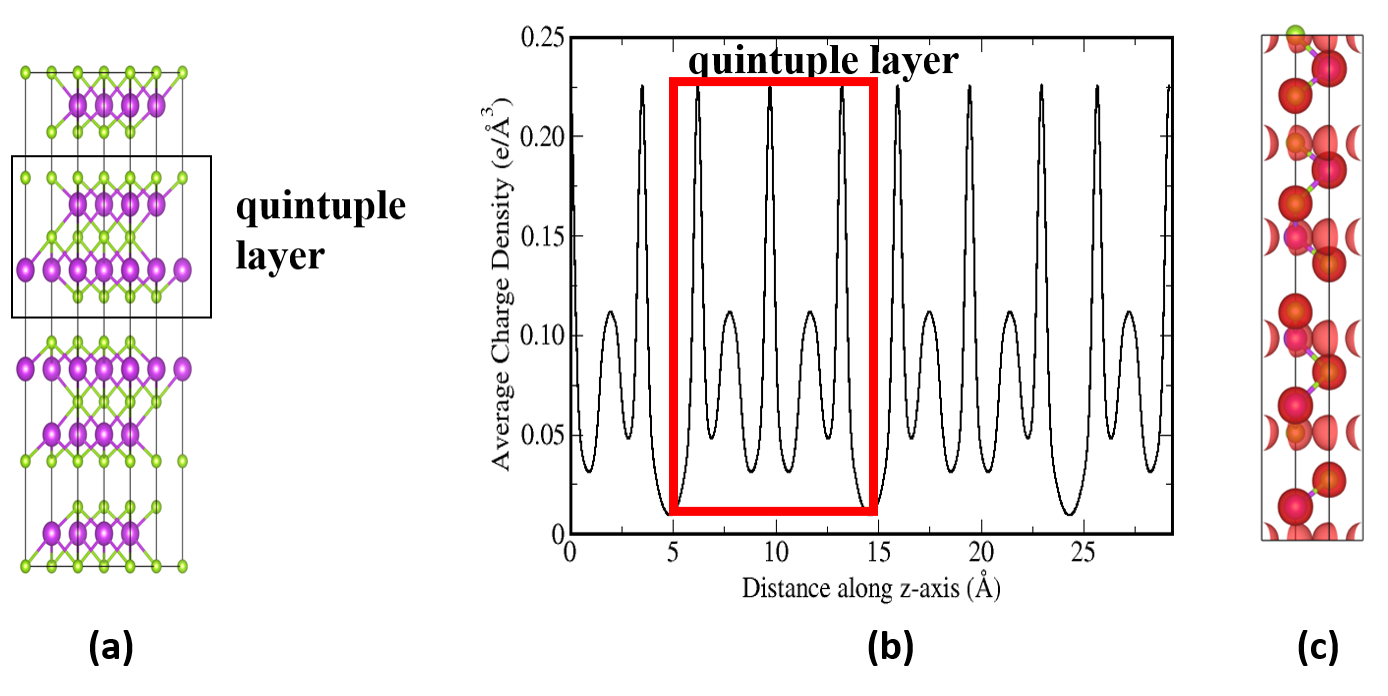}
\caption{(a) Layered structure of bulk Bi$_2$Se$_3$, (b) Average charge density on the plane of the layers vs the distance along the perpendicular to the layers ,(c) Charge density distribution within the bulk structure of Bi$_2$Se$_3$. Bi and Se are represented by pink and yellow balls, respectively.}
\label{fig:charge_km}
\end{figure}
\begin{figure}[h!]
\centering\includegraphics[width=14cm]{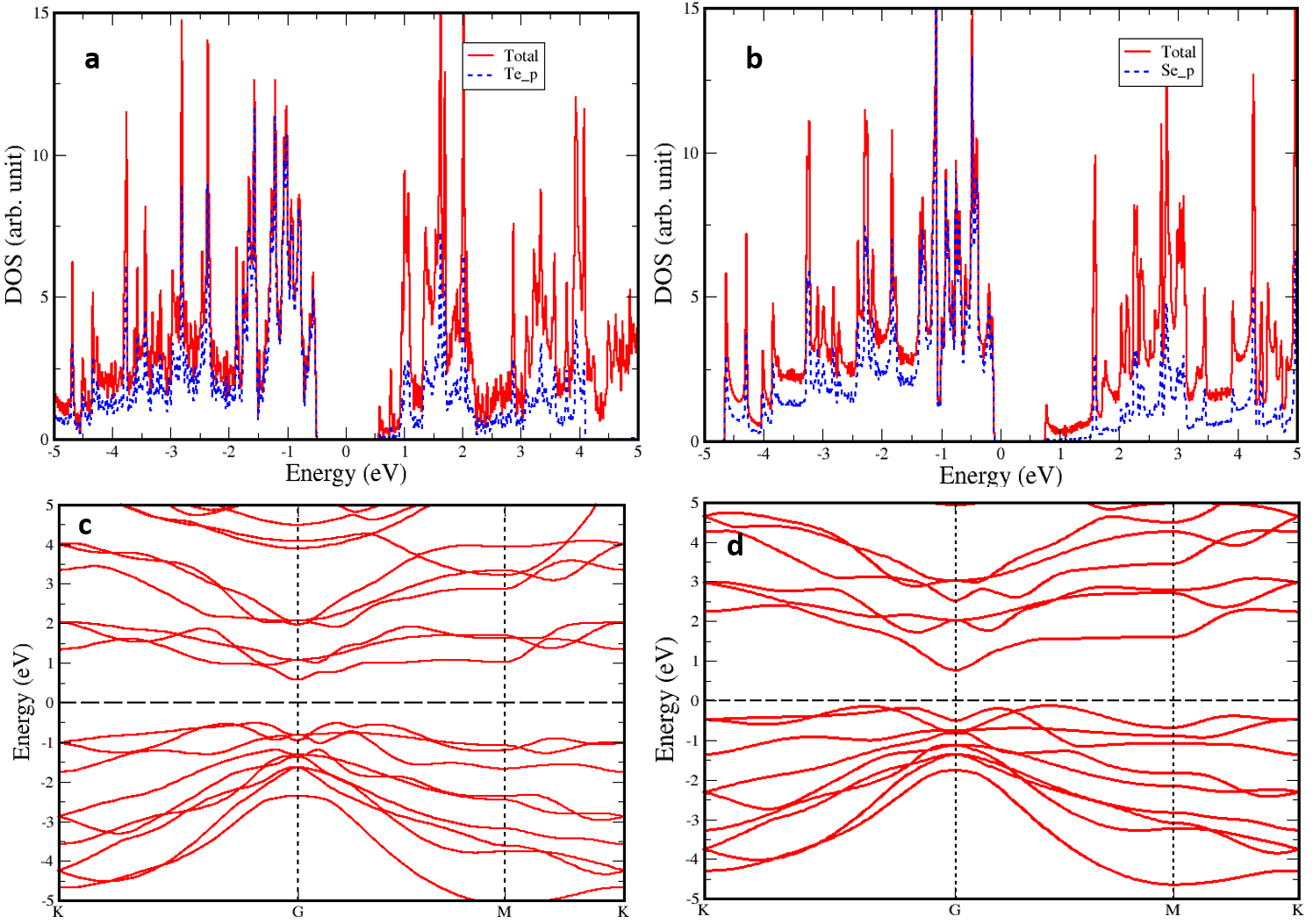}
\caption{Density of states of a quintuple layer of (a) Bi$_2$Te$_3$ and (b) Bi$_2$Se$_3$ and the corresponding band structures of (c) Bi$_2$Te$_3$ and (d) Bi$_2$Se$_3$. In (a)\&(b) Fermi level is set to zero.}
\label{fig:dos}
\end{figure}
\begin{figure}[h!]
\centering\includegraphics[width=12cm]{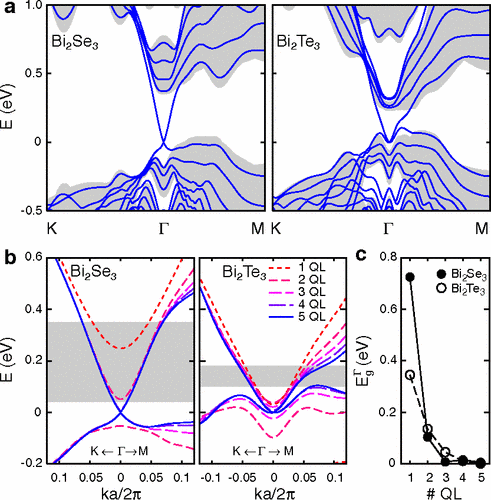}
\caption{(a) Band structure of the 5 quintuple layers slabs of Bi$_2$Se$_3$ and Bi$_2$Te$_3$ (lines) superimposed with the projected band structure of the corresponding bulk materials (shaded areas). (b) Evolution of the surface state band dispersion in the vicinity of $\Gamma$ point as a function of slab thickness. Shaded areas show the bulk band gaps. (c) Band gaps at the $\Gamma$ point induced by the interaction between the surface states as a function of slab thickness.  Reported with permission (Fig.  1, \cite{yazyev2010-prl}) 2010, APS.}
\label{fig:band_bulk}
\end{figure}
In Fig. \ref{fig:charge_km}, we have shown the layered structure of Bi$_2$Se$_3$. Similar layered structure is observed for Bi$_2$Te$_3$, Sb$_2$Se$_3$ and Sb$_2$Te$_3$. 
The existence of weak interaction between the layers is evident from the charge density distribution as shown in Fig \ref{fig:charge_km}b. 
For this analysis we adopted the computational method as discuss in our earlier study \cite{mondal2019} on similar structure of C$_2$Ti$_3$. 
Using the similar method, the lattice parameters of Bi$_2$Se$_3$ and Bi$_2$Te$_3$ topological insulators have been calculated. 
For Bi$_2$Se$_3$, the calculated values of the lattice parameters are a = b = 4.17 \AA, and c = 29.18 \AA, whereas, these lattice parameters are slightly longer for Bi$_2$Te$_3$ (a = b = 4.32 \AA and c = 30.02 \AA) as compared to those of  Bi$_2$Se$_3$. 
This small increase in lattice parameters may be attributed to the bigger atomic size of Te (2.10 \AA) than that of Se (1.20 \AA). 
We have found that the inter-layer distance between two consecutive quintuple layers in Bi$_2$Se$_3$ is 2.71 \AA, whereas, for the case of Bi$_2$Te$_3$ the quintuple layers are separated by 2.77 \AA. 
Our calculated structural parameters closely match with the experimental values\cite{betancourt2016complex}.
To analyse the bonding between the quintuple layers we have calculated the charge distribution and the average charge density on the plane of the layers and  plotted it along the axis perpendicular to the layers. The corresponding results are shown in Fig \ref{fig:charge_km} b \& c.
From these figures it is clear that the charge accumulation between the quintuple layers is negligible. Due to the weak inter-layer interaction mechanical exfoliation of one or multiple quintuple layer(s) of Y$_2$X$_3$ through sonication or Scotch tape or peeling using AFM tip is possible. Details of the synthesis process of Y$_2$X$_3$ is discussed in the next section.
The electronic structure of each quintuple layer of  Y$_2$X$_3$ can be understood from their band structure and the corresponding density of states (DOS). 
For this reason we have calculated the band structure and DOS of a single quintuple layer of Bi$_2$Se$_3$ and Bi$_2$Te$_3$ and the corresponding results are shown in Fig \ref{fig:dos}. To get more insight into the electronic structure and its nature in the multiple quintuple layers of these TIs, the band structure of 5 quintuple layers slabs of Bi$_2$Se$_3$ and Bi$_2$Te$_3$ superimposed with the projected band structure of the corresponding bulk materials has been shown in Fig. \ref{fig:band_bulk}. 
This band structure, reported in  \cite{yazyev2010-prl}, shows that the band near the Fermi level belongs to the surface states of the topological insulators.
Note that the metallic surface states of the bulk TIs are immune to the perturbation due to the back-scattering from ordinary defect sites and carry current even at high barriers which may stop other metallic states. 
Such peculiar property makes the topological insulator unique.
From our calculated DOS (Fig. \ref{dos}) it can be seen that the states near the Fermi level are dominated by the p-orbitals of Se or Te atom. 
Therefore, it is clear that the surface states are coming from the Y atoms (atoms indicated by green colored spheres in Fig. \ref{fig:charge_km}) of TI group of formula X$_2$Y$_3$. 
Here, we wish to highlight that presence of the conical shaped surface states with Dirac point (graphene like) in the band structure of TI makes it unique from other 2D materials.

For the purpose of modelocking, few layers of Y$_2$X$_3$ are attached to the fiber as SA. For example, Dou et al. \cite{dou2014mode} used 10-16 layers of Bi$_2$Se$_3$ for modelocking at $\sim$ 1 $\mu$m. 
On the other hand, thin film of 15 $\mu$m of Bi$_2$Se$_3$ has been used by Lee et al. \cite{lee2018femtosecond} to achieve modelocking at $\sim$ 2 $\mu$m.
Moreover, H Liu et al. in 2014 \cite{liu2014femtosecond} reported modelocking at 1.55 $\mu$m using nanosheets of Bi$_2$Se$_3$. 
These results show that the various thickness of Bi$_2$Se$_3$ as SA can give modelocking at different wavelengths.
It is known that the electronic structures of the films of Bi$_2$Se$_3$ depends on the thickness of the films due to the quantum confinement of the electrons.
In a theoretical study, Yazyev et al. \cite{yazyev2010-prl} reported that the band gap of Bi$_2$Se$_3$ varies with the thickness of the film.
Such variations in the band gap may be the reason for TISA based modelocking at different wavelengths with the various thickness of Bi$_2$Se$_3$ films. 
Here, we wish to mention that the calculation of the band gap of any material is very challenging and also not always comparable to the experimentally measured band gap. 
However, the relative behavior of the band gaps of various thickness of the films is more often consistent with the experimental observations.
From Fig. \ref{fig:band_bulk}, we can see that the variation in bulk band gap is wider for Bi$_2$Se$_3$ than that of the Bi$_2$Te$_3$. 
This may lead to the broad absorption spectrum for the different Bi$_2$Se$_3$ based SAs.
Moreover, the  Fig. \ref{fig:band_bulk} clearly shows that the band structure of Bi$_2$Se$_3$ has Dirac like band which is distorted in the case of  Bi$_2$Te$_3$. 
Note that the presence of Dirac cone in the band structure is mainly responsible for the broad band saturable adsorption property of graphene \cite{bhattachraya2016efficient}. 
Therefore, from the above discussion we can conclude that Bi$_2$Se$_3$ can be used for modelocking at different wavelengths (near-IR to mid-IR regime) as found in the literature, whereas, Bi$_2$Te$_3$ is found to be useful as SA for modelocking mostly at $\sim$1.55 $\mu$m.

It is highlighted in our discussion that the modelocking frequency corresponding to a SA depends on the band structure of the materials. 
Many theoretical and experimental investigations show that the band structure of a nanomaterials gets modified when its open surfaces are contaminated by external agents like oxygen, hydrogen, hydroxyl groups etc. For example, it is reported that the band gap of graphene can be modulated by reducing it with varying concentration of oxygen \cite{jin2020,bhattachraya2016efficient}. 
Moreover, it is reported that most of the 2D layered MXenes (e.g, Ti$_2$C, Ti$_3$C$_2$ etc.) become semiconductor from metal due to the presence of -O or -F groups on the surface \cite{jing2017}. 
Such changes in the band gap occur due to the removal of the surface electrons via the external groups. 
Therefore, it is important to mention that the change in the band gap of newly discovered 2D layers of TIs due to the external surface contaminating groups like -O, H, OH is a open question to the researchers. To the best of our knowledge, there is no such report exist in the literature where the above question has been addressed.
Such surface contamination may change the modelocking performance of SA, thereby may hinder the operation of the fiber lasers with SA exposed to the environment. 
We believe that this review article will motivate the experimentalists as well as theoreticians to carry out detailed investigation on the effect of environmental parameters on the performance of the fiber lasers with SA made of layered nanomaterials.

\section{Fabrication techniques of TISAs}
Topological insulators can be synthesised and fabricated as a saturable absorber using various methods. 
Some of the techniques, such as, mechanical exfoliation (ME), mechanical trituration, hydrothermal exfoliation (HTE), polyol method, etc.  \cite{sotor2014mode, sotor2014harmonically, jung2014femtosecond} are easy to process and cost effective, while other techniques, such as, molecular beam epitaxy (MBE), chemical vapour deposition (CVD), liquid phase exfoliation (LPE), self-flux method, pulsed magnetron sputtering technique, pulse laser deposition (PLD), etc. \cite{chi2014all, lee2015femtosecond, lee2016pulse, lee2018femtosecond, sobon201110} provide better uniformity and control over stacks of monolayers,
In the following, each of these processes have been discussed highlighting their importance as saturable absorber.

\subsection{Extraction using mechanical exfoliation}
\begin{figure}[h!]
\centering\includegraphics[width=12cm]{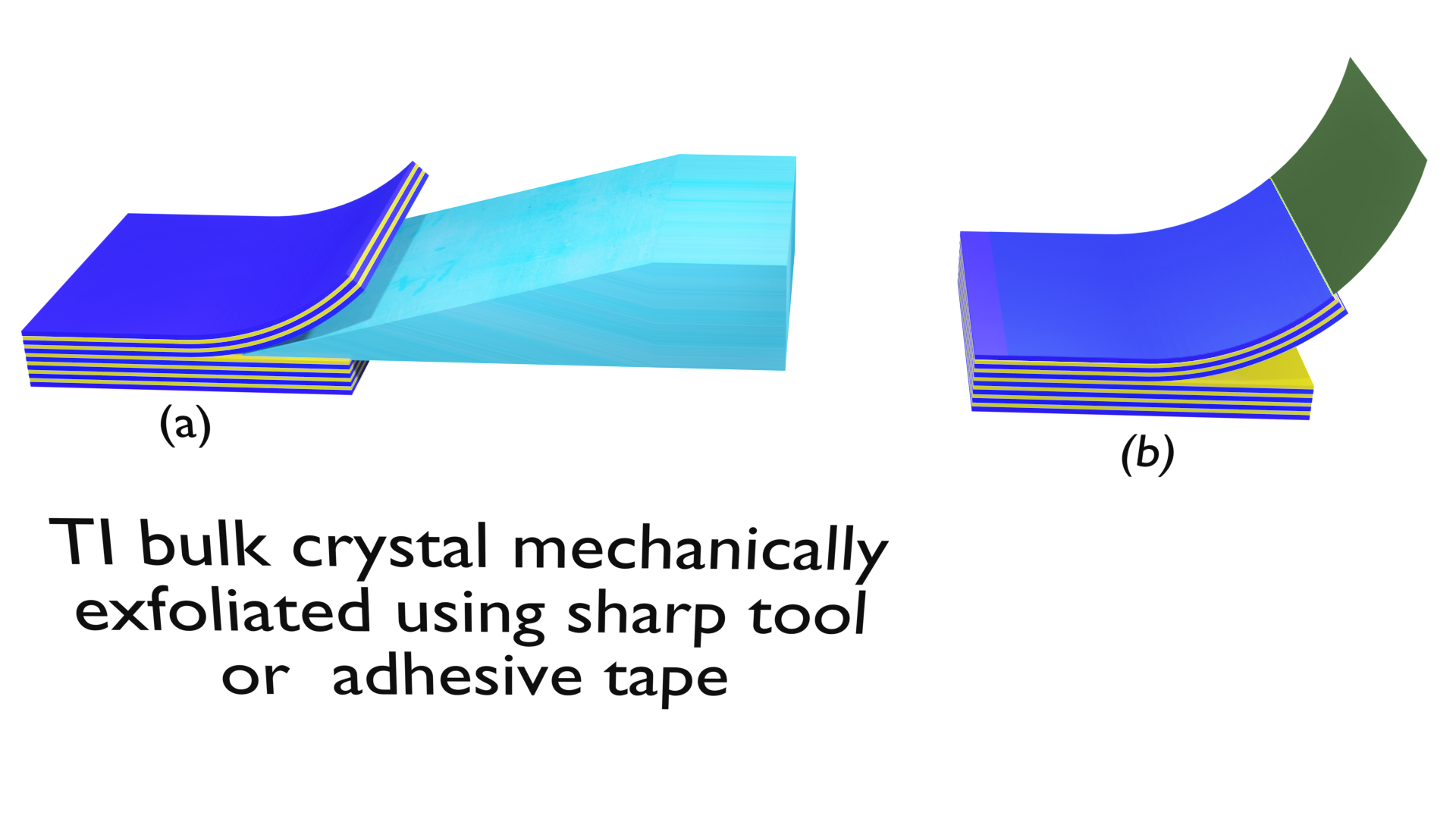}
\caption{Mechanical exfoliation setup for fabrication of TI 2D layers using (a) sharp blade and (b) scotch tape}
\label{ME}
\end{figure}

This method is the cheapest and easiest one among all other fabrication techniques.
In this technique, layered TIs can be extracted using a scotch tape or a sharp blade. 
It can be carefully transferred at the fiber end-facet to use as a saturable absorber. 
The schematic diagram of both of these processes of ME has been shown in Fig. \ref{ME}. 

This mechanism can produce several $\mu$m thick multilayered-flake TISA, which can also be attached over a side-polished fiber. 
However, manual peeling of layers compromises the repeatability of the modelocker \cite{jung2014femtosecond, lee2015femtosecond}.
An index matching oil was used on the top of polished surface to get better interaction between the light through the core of the fiber and prepared TISA. 
The interaction length is important here for optimum saturable absorption as the evanescent wave and the guided wave shares the energy of the pulse. If the interaction length is longer, the maximum amount of pulse energy will be absorbed, hence, increase the modelocking threshold and decrease the spectral bandwidth. 

\subsection{Synthesis using hydro-thermal exfoliation}
\begin{figure}[h!]
\centering\includegraphics[width=12cm]{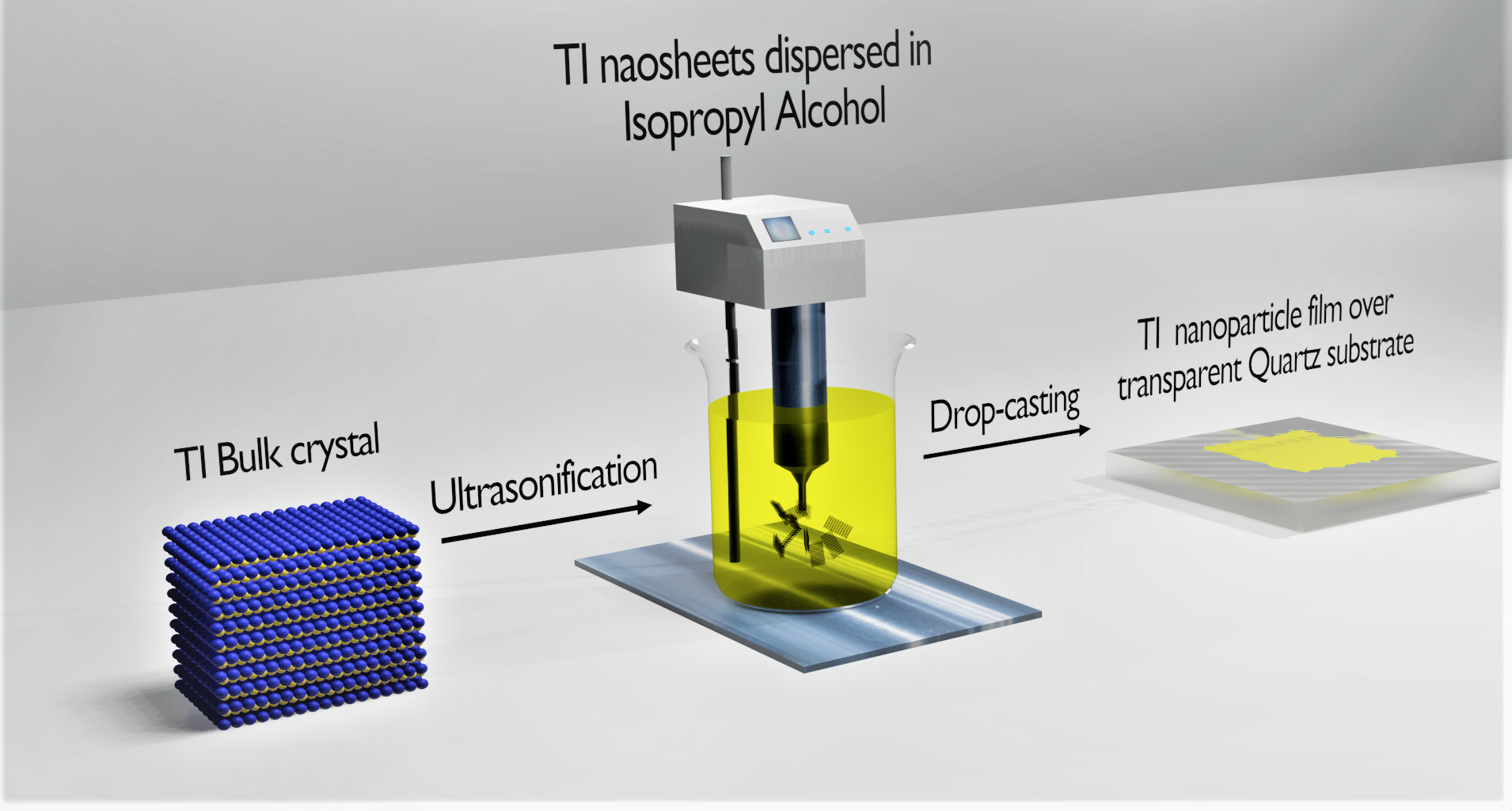}
\caption{TI nanoparticle synthesis by hydro-thermal exfoliation process}
\label{HTE}
\end{figure}

Another easy method to fabricate TISA is hydro-thermal exfoliation  (HTE) (see Fig. \ref{HTE}) or intercalation which is an optical deposition method \cite{zhao2012ultra, luo20132, liu2014dual, jin20183}.
It generally involves TI nanosheets being dispersed in an organic solvent which is followed by ultrasonication for $\sim$30 minutes. After that, the TI-solvent solution is deposited either over the fiber end-facet or on the transparent substrates like quartz, BK7, sapphire, glass slides, etc. which is then placed at the fiber end-facet of a single mode or micro-structured fiber. 
Another way is to submerge a tapered fiber in the TI solution to realize a TISA. Sometimes, the hydrothermal exfoliation is followed by drop casting and spin coating processes to fabricate TISA. 
To ensure the high quality deposition, during the whole process, an ASE source (at one end) and photo-detector (at the other end) should be connected and monitored the escaping of light.

\subsection{Synthesis by liquid phase exfoliation}
\begin{figure}[h!]
\centering\includegraphics[width=12cm]{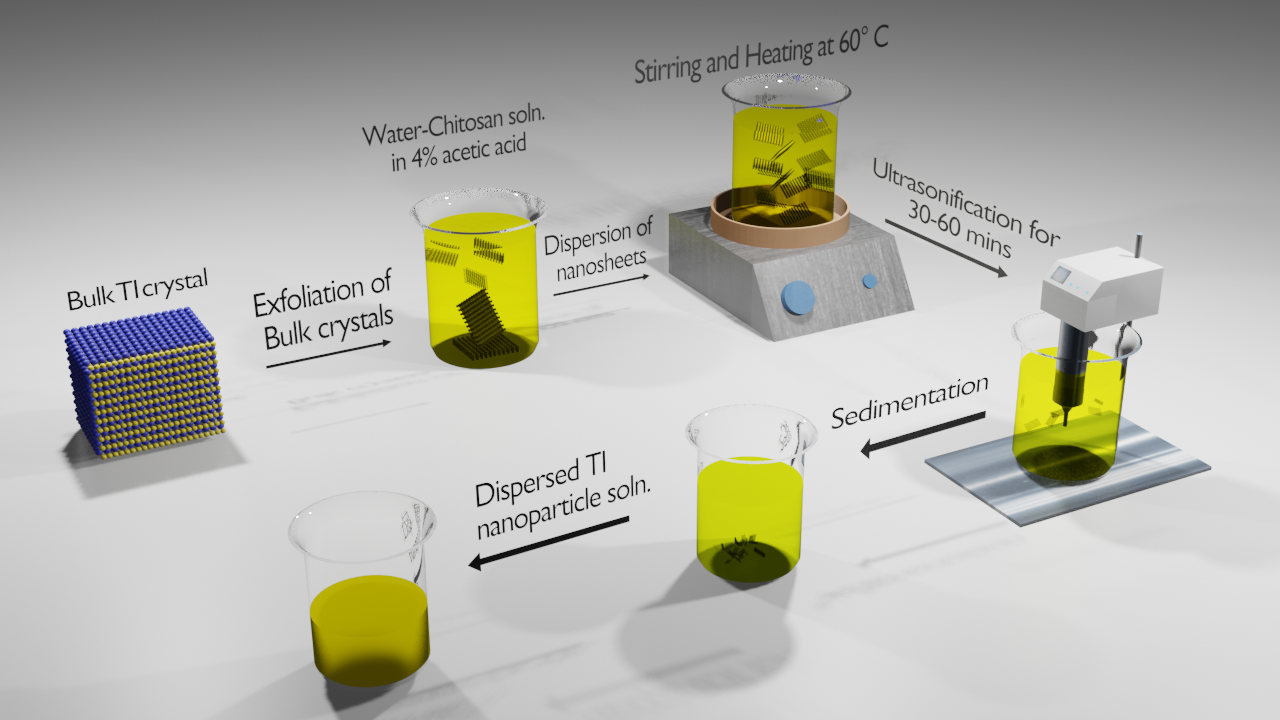}
\caption{Synthesis of TI nanoparticles by liquid phase exfoliation process}
\label{LPE}
\end{figure}

Liquid phase exfoliation is a type of wet chemical synthesis process, widely used for 2D nanomaterials, known for its simplicity, cost-effectiveness and good repeatability \cite{boguslawski2014mode,guo2015observation, guo2015topological, liu2020ultrathin}.
This synthesis process is shown in Fig. \ref{LPE}. The preparation involves taking few milligrams of bulk TI crystal in water-chitosan solution dissolved in 4$\%$ acetic acid. 
The resultant solution is heated ($\sim$60\degree C) and stirred simultaneously, and sonicated for $\sim$60 minutes. 
The mixture is left for $\sim$24 hours to separate the non-exfoliated, thick layers by sedimentation.
The final solution droplets are deposited over a side-polished fiber to prepare the SA as shown in Fig \ref{LPE}. 
The realization of TI nanoparticles can also be derived from corresponding compound powders by hydrothermal process and then adding that mixed powder to N-methyl-2-pyrrolidone solution, followed by ultrasonication process to form well dispersed solution. 
After that, polyvinyl alcohol (PVA) powder is mixed with the resultant solution along with ultrasonic agitation to prepare the PVA-TI suspension. 
This solution is placed onto a transparent substrate by spin-coating and finally oven dried to fabricate the PVA-TI film \cite{guo2015observation}. 
Tellurene, a newly discovered TI, composed of stacked layers of tellurium atoms arranged in a trigonal crystal lattice, can also be fabricated using LPE. Tellurium bulk crystals are granulated to powder form and further grinned in pure ethanol for dispersion. 
For better dispersion, the nanosheets undergo $\sim$2 hours of sonication process and finally pass through centrifugation to eliminate the larger particles from the desired TI nanoparticles \cite{liu2020ultrathin}.

\subsection{Film fabrication by polyol method}
Another traditional fabrication technique is polyol method which forms thin film of PVA-TI composite, placed between two fiber ferrules in a laser resonator \cite{zhao2012wavelength, liu2014femtosecond, luo2014observation,dou2014mode,li2015band}. 
General process involves preparing a solution by ultrasonicating TI nanosheets in acetone. 
It is further mixed with few milliliters of aqueous solution of PVA and once again undergoes ultrasonication process for around half an hour. 
Finally, the solution is placed over a small transparent substrate and evaporated to form TI-PVA film.
This film can be used as a TISA as inserted between two optical fiber connectors \cite{liu2014femtosecond}.
Another alternative way to realize the polyol method is to mix the pure and dispersed TI solution in deionized water, followed by repeated centrifugation and ultrasonication. 
To prepare the film, TI water solution is added dropwise to deionized water-soaked filter paper and dried in heater. 
At the fiber end-facet the filter paper is placed, which is finally removed by applying acetone solution repeatedly to leave behind only the TI film \cite{dou2014mode}. 
Moreover, another type of polyol technique is to take compound materials of TI along with polyvinyl pyrrolidone and ethylene glycol, which are added to a flask comprising of Teflon coated magnetic stirrer.
This flask is attached with a reflux condenser and put over a furnace at $\sim$190$\degree$ C temperature with simultaneous constant stirring for couple of hours. 
The resultant solution is cooled, centrifuged and cleaned with isopropyl alcohol and dried at $\sim$60\degree C on a transparent substrate to achieve the TISA film \cite{zhao2012wavelength}.

Apart from these solution based chemically synthesised processes of preparing TI nanoparticle films, people have also tried some less known fabrication process such as, Solution-phase exfoliation. 
It is another easy fabrication technique whereby, bulk TI powders are added to chitosan acetic solution. 
After sonication process for around 30 hours, the particles get evenly mixed in the solvents. 
This solution is further mixed with aqueous PVA solution and is ultrasonicated for $\sim$30 mins. 
The final solution is evaporated in an oven over a glass slide to develop TI: PVA film \cite{guo2015dual}.

\subsection{Growth of TI nanoparticle film by chemical vapour deposition}
A common technique for growing TI nanoparticles in bottom-up method is Chemical Vapour Deposition (CVD)  \cite{wei2019large,guo2019output}. CVD provides advantage over methods such as ME, HTE and LPE in terms of uniformity of monolayers, control over number of layers, higher damage threshold, etc. In this method, SiO$_2$ substrate is cleaned with acetone, ethanol and de-ionized water beforehand is placed inside the CVD tube. Bi$_2$Te$_3$ powder is kept at a particular distance (typically $\sim$22 cm) away from the substrate in the temperature controlled horizontal tube furnace. Under vacuum condition, the tube is heated up to 350\degree C at a uniform rate in the presence of pure Argon gas. The growth of Bi$_2$Te$_3$ over the substrate took $\sim$30 minutes, after which the tube is rapidly cooled down with a blower. A pyrolysis tape is used to remove the thin layer of TI from the SiO$_2$ substrate. Then the fiber connector is placed directly over the TI layer contained pyrolysis tape and is heated for 10 minutes over 130\degree C to vaporize the tape material. By this way, TI layer is transferred permanently over the end-facet of the fiber connector.

\begin{figure}[h!]
\centering\includegraphics[width=12cm]{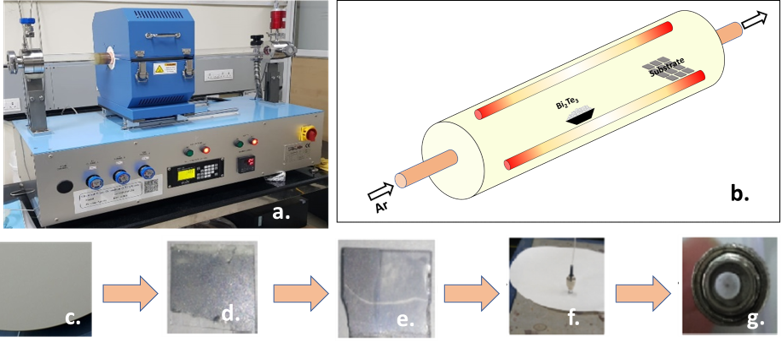}
\caption{(a) CVD setup for fabrication of Bi$_2$Te$_3$ thin films. (b) Stages of preparation of Bi$_2$Te$_3$ SA.}
\label{CVD}
\end{figure}

\begin{figure}[h!]
\centering\includegraphics[width=12cm]{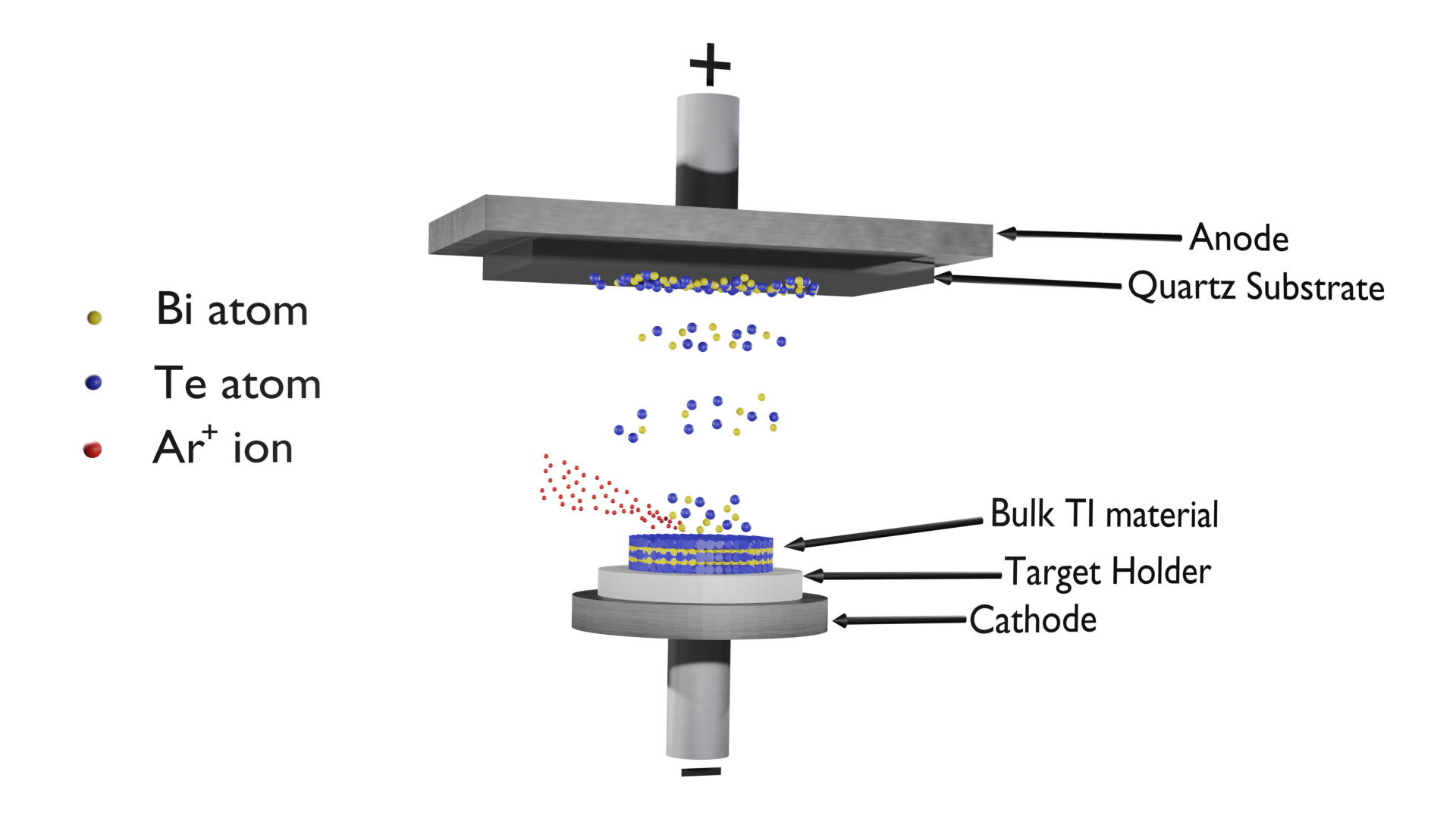}
\caption{DC-pulsed magnetron sputtering technique for TI film development}
\label{PMS}
\end{figure}

\subsection{TI coating using pulsed magnetron sputtering}
Pulsed magnetron sputtering can also be applied to deposit TI nanoparticles over quartz substrate, side-polished fiber and connector end-facet fiber  \cite{boguslawski2015dissipative}. 
This technique (Fig. \ref{PMS}) involves having high purity antimony and tellurium powders for preparing a stoichiometric mixture which is grounded and homogenized. Under vacuum condition and $\sim$680\degree C, the mixture is placed in quartz ampoules covered with thin pyrolytic carbon layer, where it takes an hour for synthesis. A constant rotation of the quartz ampoule is needed for proper mixing of the alloy. After cooling down the ampoules to room temperature, the resultant mixture is powdered in a grinding machine and fritted in a cast to fabricate cathodes for magnetron sputtering. The TI layers are deposited over side-polished fiber through a mask ($\sim$100 microns to few millimeters) in single planar magnetron. Here, the mask can uniformly control the deposited length of TI nano-coating on the fiber.  

\subsection{TI film preparation using self-flux method}
Self-flux method is another technique to fabricate TISA. This technique is generally used to develop AuTe$_2$Se$_{4/3}$ based SA. Initially high purity of Au, Te and Se powders are prepared. Then, 2-4 gms of each powder are taken in evacuated silica tube with a vacuum of $\sim$10$^{-5}$ mbar and heated at 800\degree C for $\sim$10 hours in a furnace. The next step is to decrease the temperature very slowly and uniformly to 450\degree C. The AuTe$_2$Se$_{4/3}$ nanoparticle of 60 mg is dissolved in 20 ml ethanol, followed by ultrasonication for even dispersion in ethanol. Finally, $\sim$40 $\mu$L of the solution is dropped over the fiber end-facet and dried at room temperature \cite{liu2020ultrafast}.

\begin{figure}[h!]
\centering\includegraphics[width=12cm]{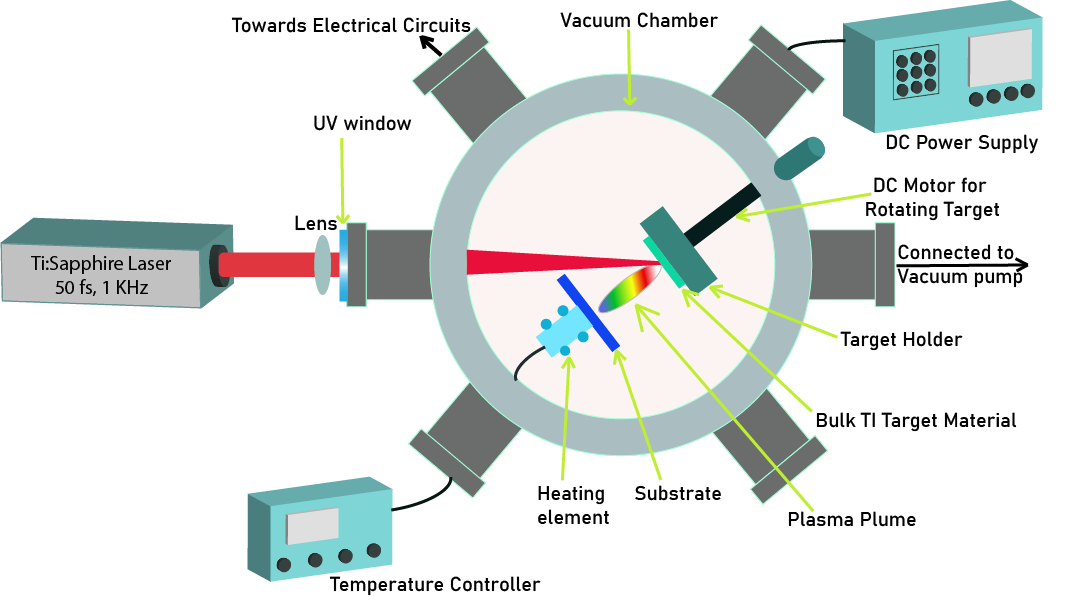}
\caption{Pulsed laser deposition technique for TI film development}
\label{PLD}
\end{figure}

\subsection{TI film preparation using pulsed laser deposition}
Pulsed Laser Deposition (PLD) technique has been proved to be the most efficient and simple one to grow 2D nanomaterials \cite{li2015band, yan20152,chen2016high,liu201670, li2017high}. It does not require high growth temperature or extremely high vacuum environment. Moreover, constant composition during fabrication, excellent repeatability, uniformity, longer interaction length, higher power tolerance, robust TI film, no need for index matching gel and simple fabrication process are among the other noticeable advantages of PLD technique. Also, PLD lowers the non-saturable losses and provides good contact with the fiber. The number of layers too, can be easily controlled using deposition parameters such as laser pulse energy, temperature and deposition time. Various pulsed lasers such as Q-switched nanosecond Nd:YAG laser, Nd:YVO$_4$ picosecond laser, Kerr lens modelocked Ti:sapphire laser, etc. can be used for this purpose. Peak power is usually in the order of kW to MW and the vacuum chamber pressure is around 10$^{-4}$ to 5 x 10$^{-4}$ Pa. The time required for deposition over the fiber is between 60 to 90 minutes for uniform deposition.

In pulse laser deposition, the high energy laser pulses help to melt, evaporate and ionize the TI material from the bulk TI. This process is called ‘laser ablation’ where an extremely bright plasma plume (consisting of neutrals, ions, and electrons) expands very fast from the bulk crystal. The velocity of the plume can reach up to ~104 m/s. There are many functions, such as, laser intensity, background gas pressure, absorption coefficient of target material, plasma density, distance between target to substrate, temperature of the substrate, etc. can govern the growth of the film \cite{chrisey1994pulsed}. The ablated TI ions then gather on a suitably placed substrate (in this case, fiber facet or side-polished or tapered fiber surfaces) where it condenses slowly to develop a thin film. However, all these optimizations are limited to a set of target and substrate materials, hence optimized conditions of one material cannot fit into another.

People have also tried Mechanical Trituration technique, which is a very common industrial technique to prepare the amalgams, to get TI particles in the order of sub-microns. It is possible to fabricate TISA of not less than $\sim$500 nm thick nanosheets by mechanical polishing of doped TI bulk crystals in a commercial triturator  \cite{lin2014soliton, lin2015using}.

\begin{figure}[h!]
\centering\includegraphics[width=12cm]{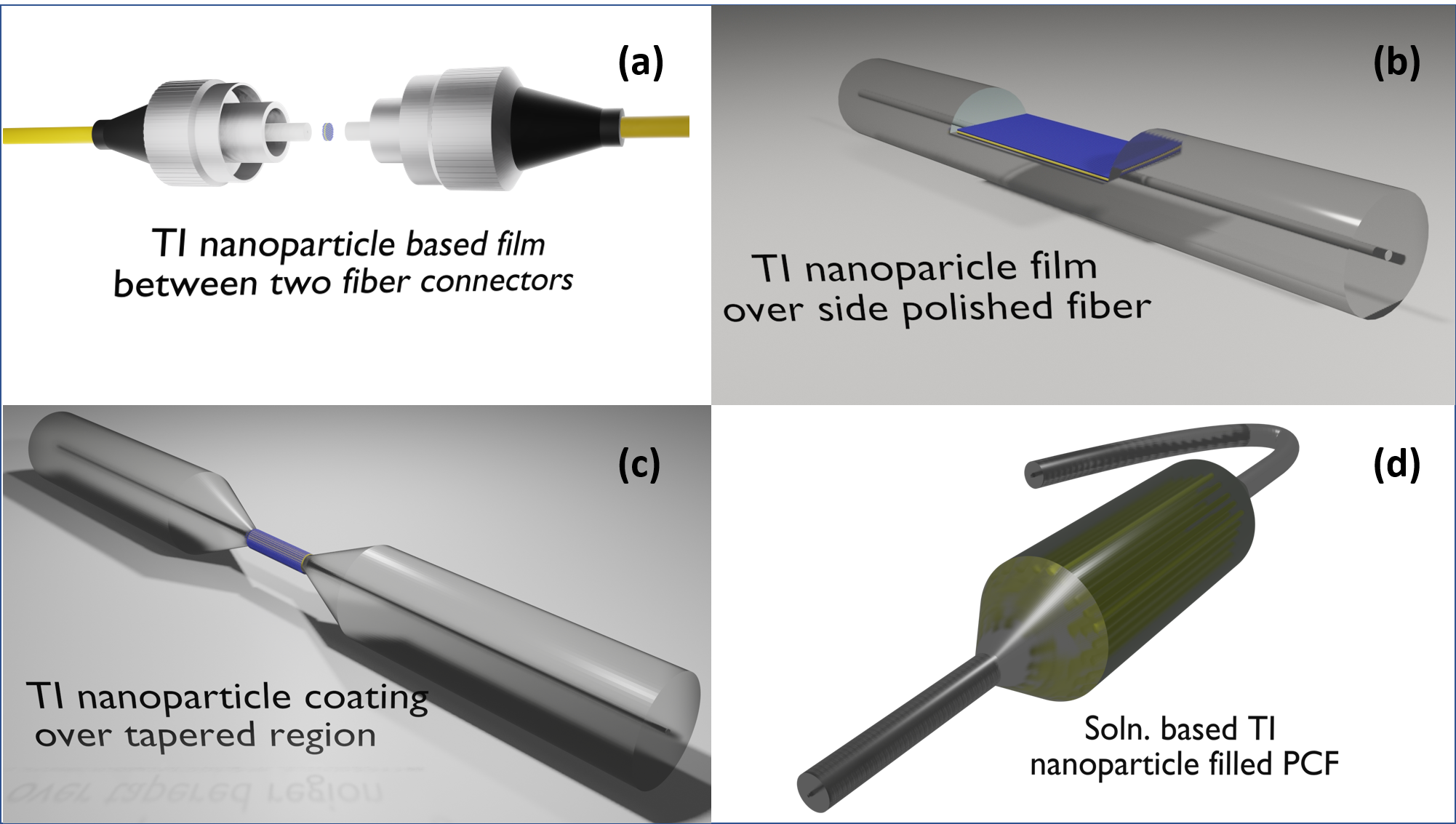}
\caption{Device integration processes of TISA: (a) between two fiber connectors or end-facets/ferrules, (b) TI coating over side polished fiber, (c) TI coating over tapered fiber region and (d) photonic crystal fiber having clad-filled solution based TI nanoparticles.}
\label{integration}
\end{figure}

\subsection{TI coating in fibers using optical deposition technique}
Another type of fabrication process is Optical Deposition (OD) technique where a light source is used during the deposition process over the fiber sample.
To use this method for preparation of Bi$_2$Te$_3$ based SA, a stoichiometric ratio of bismuth chloride (BiCl$_3$) and sodium selenide (Na$_2$TeO$_3$) are mixed together with the aid of moderate ethylene glycol with continuous stirring. 
The dissolved mixture is put into an autoclave to produce grey Bi$_2$Te$_3$ powder. The powder is washed with distilled water and ethanol and finally dried at 60\degree C in vacuum. 
Dry powder is again mixed in ethanol solution having a concentration of typically $\sim$100 $\mu$g/mL. 
The dispersed material is finally optically deposited over tapered SMF fiber \cite{yin2015soliton}. 

The optical deposition technique is a monitoring process of TI materials over the optical fiber. It uses an illuminated light (IL) source for the deposition of TI solution over the fiber. A power meter is usually attached to monitor the process. Initially the coupled light from IL escapes from the core of the optical fiber via evanescent waves. Therefore, very low light reaches the power meter. The moment TI material starts deposition, the power in the power meter increases and reaches a certain value and saturates.

\begin{figure}[h!]
\centering\includegraphics[width=10cm]{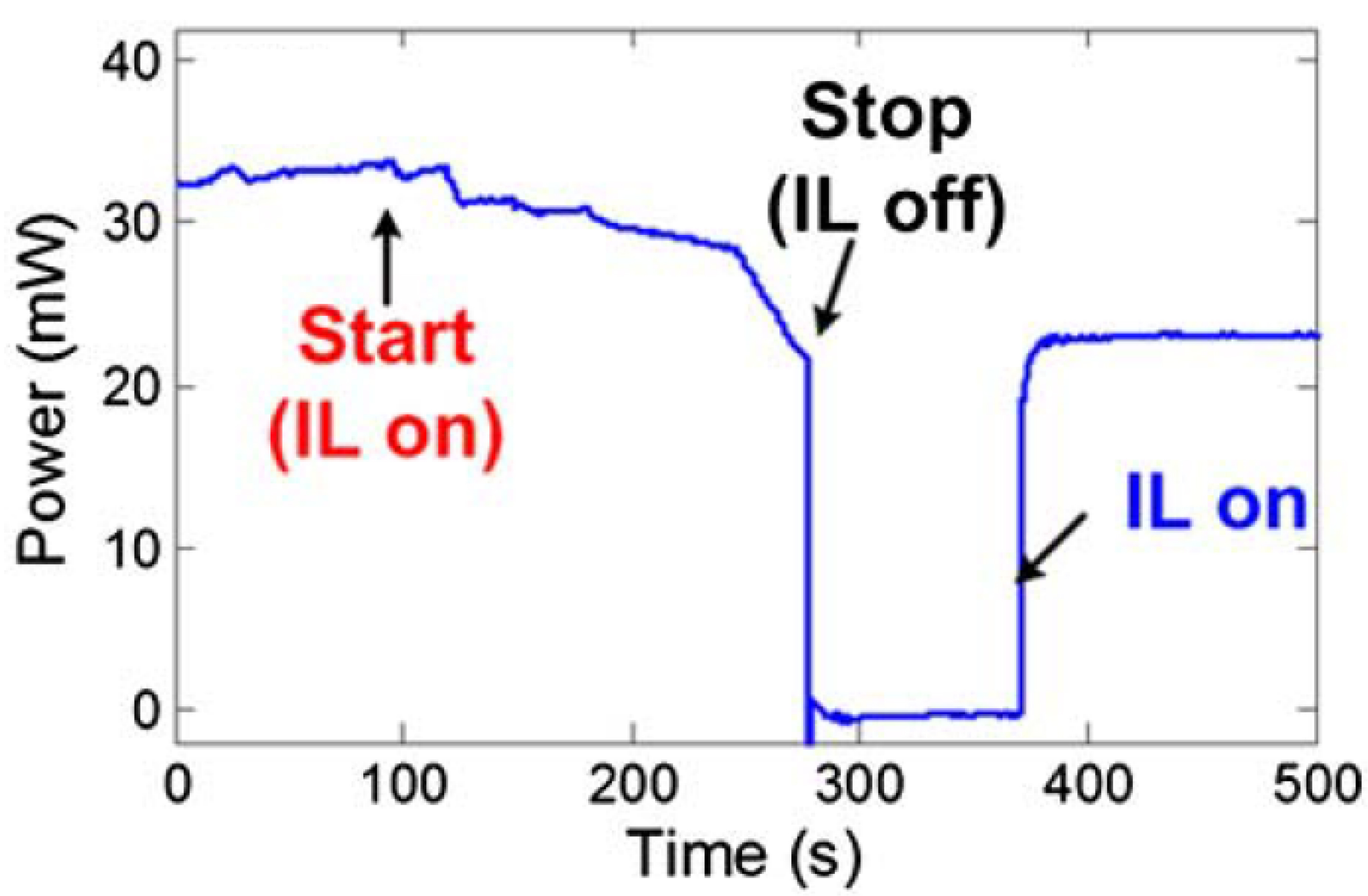}
\caption{Deposition Process in presence of IL}
\label{fig:opticaldeposition}
\end{figure}

In Fig.\ref{fig:opticaldeposition} it can be seen that the illuminated light source remains ON for the entire duration of the deposition process, after which it is turned OFF to allow the evaporation of the solvent. Again, the light can be switched ON to measure the insertion loss of the fabricated TISA.

\begin{itemize}
 \item \textit{{Sonochemical exfoliation with optical deposition technique}} 
\end{itemize}
Sonochemical exfoliation is another SA fabrication method which is very useful to prepare coating of TI nanoparticles in tapered region of optical fiber in  association with OD method. This process can be used to realize bismuthene based SA \cite{guo2018sub}. Here, Bismuth is grinded into powder and added to ethanol solution in a bottle. Few microliters of bismuth ethanol and ethanol solution are placed inside a glass bottle and kept under ice-bath and probe sonication for $\sim$15 hours, respectively. The resulting suspension goes under centrifugation process at $\sim$7000 rpm, which develops the bismuthene dispersion. Optical deposition method is used over tapered or side-polished fiber to fabricate final form of TISA.

\subsection{Deposition of TI nanoparticles in PCF using total reflection method}
Photonic crystal fiber (PCF) is one of the micro-structured fiber where solution-based TI can be inserted in the micro-holes and that can be used as a SA in a fiber laser cavity. Bi$_2$X$_3$ (where, X= Te/Se) crystal can be fabricated from bismuth oxide powder and selenium/tellurium powder and then sonicated in N-methyl 1-2-pyrrolidone solution to achieve a few nanometers thick Bi$_2$X$_3$ nanosheets solution. PCF has air holes of tens of microns radius, through which the TI solution enters. The solution is then evaporated at $\sim$40\degree C to leave behind only the TI nanosheets deposited inside the PCF \cite{gao2015stable}. Both ends of PCF can be fused and spliced with SMFs to be used as SA.

\subsection{Device integration of TISAs}
As mentioned earlier, apart from the various synthesis techniques for TI nanoparticles, in summary, there are four major processes to integrate it in fiber lasers (Fig. \ref{integration}). It can be classified by the position of integration, mainly at end-facet/ferrule,  side-polished fiber, tapered fiber and photonic crystal fiber. Deposition of SA over the fiber end-facet or ferrule is the most common and simple method as it is shown in Fig. \ref{integration}(a) \cite{zhao2012ultra, lin2014soliton, sotor2014mode,sotor2014harmonically, dou2014mode, chen2014formation,luo2014observation,liu2014femtosecond,li2015band,guo2015dual,guo2015topological,guo2015observation,lin2015using,chen2016high,xu2017bilayer,wei2019large,liu2020ultrafast}. The TI nanoparticles can be transferred in different ways, as mentioned in the previous section, however, transferring is always little bit tricky as it leaves the residue of other materials with the TI film. Therefore, it adds impurity and non-uniformity while integrating at the cross-section of the fiber, hence, degrades the performance of the SA. The saturable absorption can be controlled by the thickness of the TI layers which is limited due to the restriction on the number of layers possible on the end-facet cross-section. Moreover, this scheme does not provide high damage threshold and may enable Q-switching if the deposition is faulty. 

TISA based on side-polished fiber (Fig. \ref{integration}(b)) gives length-dependent controllability and high damage threshold than the previous one but has a large polarization-dependent loss which generates high nonlinear polarization rotation effect and also requires the SA setup to be protected from environment to avoid weakening of saturable absorption over time  \cite{sotor2014sub,jung2014femtosecond,boguslawski2014mode,chi2014all,lee2015femtosecond,boguslawski2015dissipative,lee2016pulse,lee2018femtosecond}. 

Tapered fiber scheme for integration of TISA (Fig. \ref{integration}(c)) in fiber laser resonators is also similar to side-polished one in terms of damage threshold and protection requirement of the SA. However, due to complex geometry of tapering, the deposition of TI nanoparticles is not uniform over the whole length of the TISA. Hence, linear loss of the cavity will be enhanced due to scattering at the tapered section. Therefore, it is not an absolute accurate technique since it engages complex light deposition methods  \cite{luo20132,liu2014dual,li2015band,yin2015soliton,yan20152,li2016vector,liu201670,li2017high,jin20183,guo2018sub,liu2020ultrathin}. 

Photonic Crystal Fiber (PCF) based TISA (Fig. \ref{integration}(d)) is the most stable one in comparison to others as SA materials stays in closed environment inside the air-holes in PCF, spliced directly to the SMF. 
It allows to tune the interaction intensity with the TISA by choosing different sizes of air-hole-filled SA materials in the PCF, hence the saturable absorber interaction length and the laser pulse duration change subsequently   \cite{gao2015stable}. However, PCF based TISA also has several disadvantages. It has poor performance in case of poring of nanosheets having large diameter ($\sim$20 nm) inside the longer PCF (few centimeters) even if siphon technique is adopted to fill it. Other limitations of this PCF-TISA are slow evaporation of liquids, distortion of guiding mode, stability issues in mode-locking operations, etc. These types of PCF-TISA help to control the nonlinearity and dispersion inside the fiber laser cavity, hence dissipative soliton modelocked pulses can be achieved. Due to the complexity of PCF-TISA fabrication, it is not popular to the ultrafast fiber laser community. 

\begin{figure}[h!]
\centering\includegraphics[width=18cm]{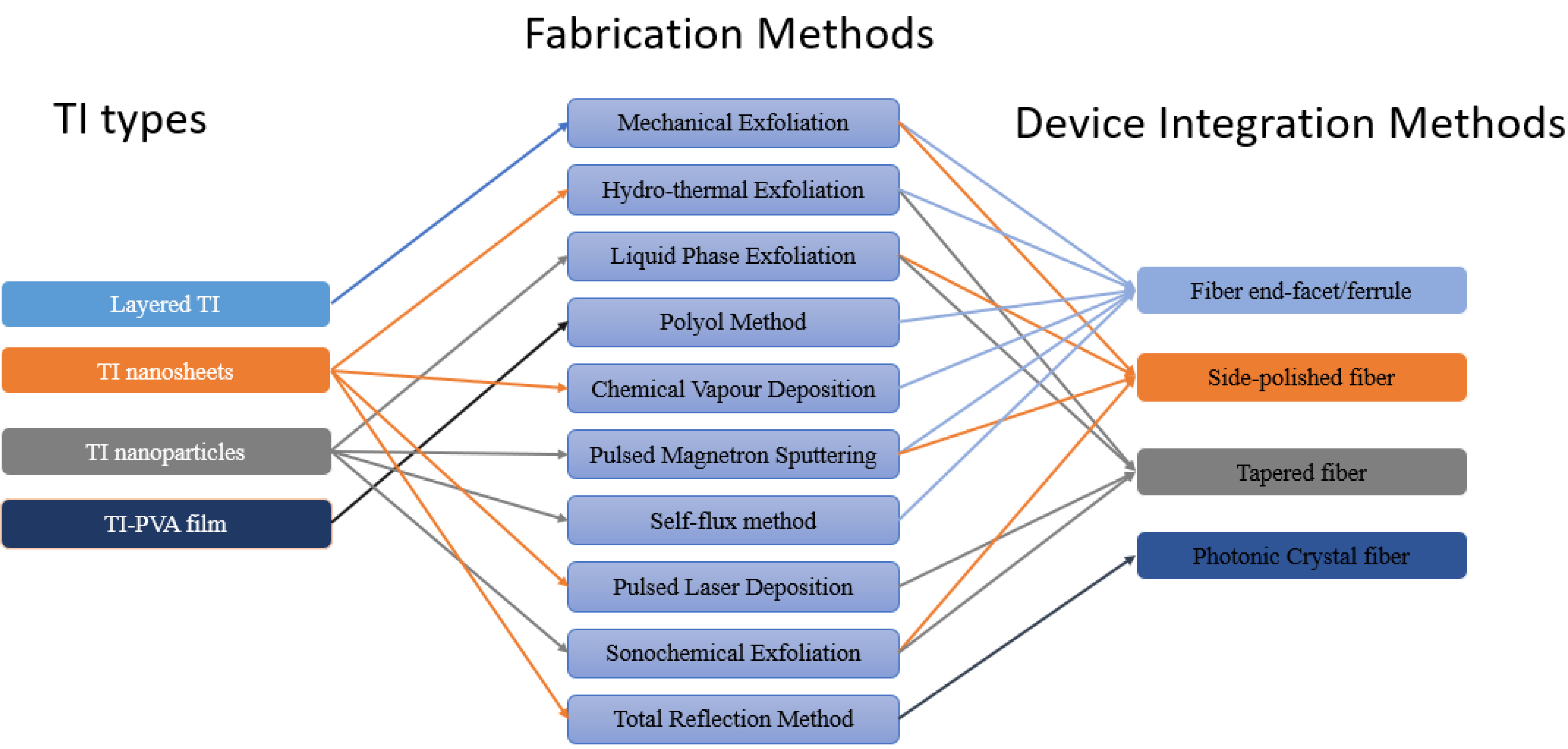}
\caption{Types of TIs and Device Integration Methods associated with Fabrication Techniques}
\label{fig:fabrication}
\end{figure}

In Fig.\ref{fig:fabrication}, the relationship among the types of TIs, device integration techniques and various fabrication methods has been shown.
The major optical characteristics affected by the type of fabrication method chosen for a particular TISA has been listed in Table \ref{table1}. The three exfoliation techniques (ME, HTE and LPE) and polyol method have poor repeatability and prone to external influences such as impurities, trapped states, etc. Thus, the optical parameters such as modulation depth, pulsewidth, etc. vary over a long range. Hence, it is difficult to predict the performance of TISA fabricated by these methods. Other fabrication methods such as PMS, CVD, OD and PLD can grow the TI material in a controlled environment, hence, have better quality of the TI thin film.

Among all the fabrication techniques, PLD can generate the shortest pulsewidth due to better control of growth of atomic layers. This happens due to low growth temperature, high vacuum growth, slow growth rate, defect-free crystalline growth, etc., makes this process incomparable.

\begin{table}[t]
\caption{Major Optical Properties of TISAs co-related with Fabrication Methods}
\centering
\arrayrulecolor{black}
\begin{tabular}{|p{1.6cm}|p{1.4cm}| p{1.6cm}|p{1.7cm}|p{1.6cm}|p{3cm}|}

\hline
Fabrication Technique & TI -Materials & Range of -Modulation    Depth & Orders of Pulsewidth (ps) & Time-Bandwidth Product & Reference \\

\hline

ME & Bi$_2$Te$_3$, Bi$_2$Se$_3$, Sb$_2$Te$_3$ & 1.8 - 20.6 & 0.128 - 12800 & 0.315 - 8836.3 &  \cite{lin2014soliton}, \cite{chi2014all}, \cite{lee2016pulse}, \cite{lee2018femtosecond}, \cite{sotor2014harmonically}, \cite{sotor2014mode}, \cite{sotor2014sub}, \cite{lee2015femtosecond}, \cite{jung2014femtosecond} 

\\
\hline

HTE & Bi$_2$Te$_3$, Bi$_2$Se$_3$ & 1.7 - 95.3 & 1.21 - 4.09 & 0.32 - 0.6843 & \cite{chen2014formation}, \cite{zhao2012ultra}, \cite{luo20132}, \cite{liu2014dual}, \cite{meng2015high} \\

\hline

LPE & Sb$_2$Te$_3$, Tellurene, Bi$_2$Se$_3$ & 3.8 - 35.64 & 1.03 - 25160 & 0.322 - 773.3 & \cite{boguslawski2014mode}, \cite{liu2020ultrathin}, \cite{guo2015dual}, \cite{guo2015observation}, \cite{guo2015topological} \\

\hline

Polyol Method & Bi$_2$Te$_3$, Bi$_2$Se$_3$ & 2.4 - 12 & 0.36 - 46 & 0.325 - 32.4125 & \cite{zhao2012wavelength}, \cite{liu2014femtosecond}, \cite{li2015band}, \cite{dou2014mode}, \cite{li2017analysis}, \cite{luo2014observation} \\

\hline

CVD & Bi$_2$Te$_3$ & 10.8 & 3220 & 674.58 & \cite{wei2019large}\\

\hline

PMS & Sb$_2$Te$_3$ & 1.5 - 5.3 & 0.167 - 1.61 & 0.4568 - 0.7 & \cite{boguslawski2015dissipative}, \cite{wang2019generation} \\

\hline

PLD & Bi$_2$Te$_3$, Sb$_2$Te$_3$ & 6.2 - 31 & 0.07 - 5470 & 0.323 - 3250 & \cite{li2017high}, \cite{lu2015yb}, \cite{chen2016high}, \cite{liu201670}, \cite{yan20152}, \cite{jin20183} \\

\hline

OD & Bi$_2$Te$_3$, Bi$_2$Se$_3$, Bismuthene & 2.3 - 9.8 & 0.193 - 7.564 & 0.342 - 0.607 & \cite{li2016vector}, \cite{yin2015soliton}, \cite{gao2015stable}, \cite{guo2018sub} \\

\hline
\end{tabular}
\label{table1}
\end{table}

\section{Characterization of TISA}
\subsection{Techniques of check surface roughness and film quality}
After fabrication of TISA, the device undergoes several surface and quality characterization to check for any defects and overall standard. $\mu$-Raman spectroscopy is used frequently to plot the shift in wavenumber vs the arbitrary intensity,  generating multiple peaks, which provide information about any defect in crystal lattice and number of layers of nanosheets \cite{shahil2012micro}. 
Using first principle method, $\mu$-Raman spectra of the materials can be calculated and by comparing the experimental and simulated peak positions, the quality of the sample and number of layers can be confirmed.
Also, atomic force microscopy (AFM) can be employed to measure the number of quintuple layers present in each TI nanosheet layer. On the other hand, scanning electron microscopy (SEM) shows microscopic images of the sample, which detects the presence of any moisture, polymer residues, grain boundary, surface uniformity and localization of the material. Energy dispersive x-ray spectroscopy (EDS) along with SEM gives us elemental information about the sample. Transmission electron microscopy (TEM) can also be done to acquire the inner stucture of the TISA such as crystal structure, morphology, etc. Moreover, X-ray powder diffraction (XRD) and X-ray photoelectron spectroscopy (XPS) are conducted to check the molecular dimensions and elemental composition of the TI nanosheets respectively.
\begin{figure}[h!]
\centering\includegraphics[width=15cm]{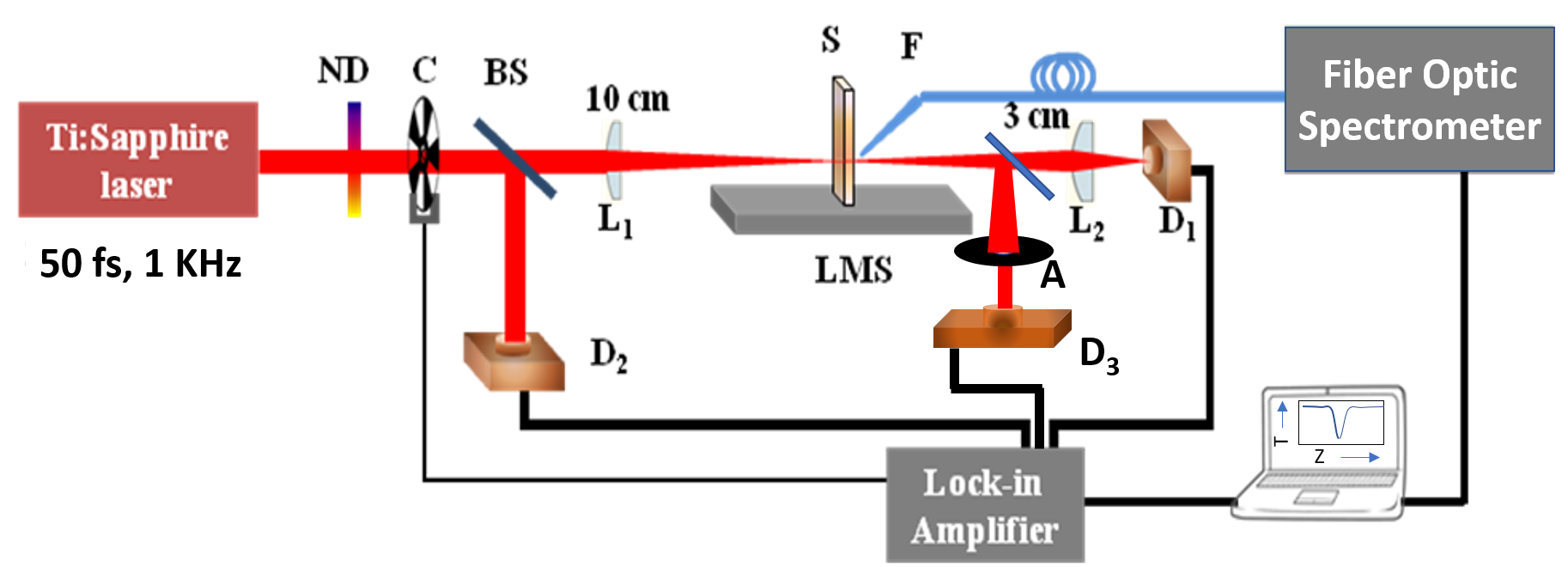}
\caption{Schematic diagram of nonlinear optical measurement - Z-scan setup; ND: Neutral density filter, C: chopper, BS: beam splitter, D: Photo-detector, L: Lens, S: sample (here, Quartz substrate having TI nanoparticle film), LMS: Linear Motion Stage and A: aperture}
\label{z-scan}
\end{figure}

\begin{figure}[h!]
\centering\includegraphics[width=16cm]{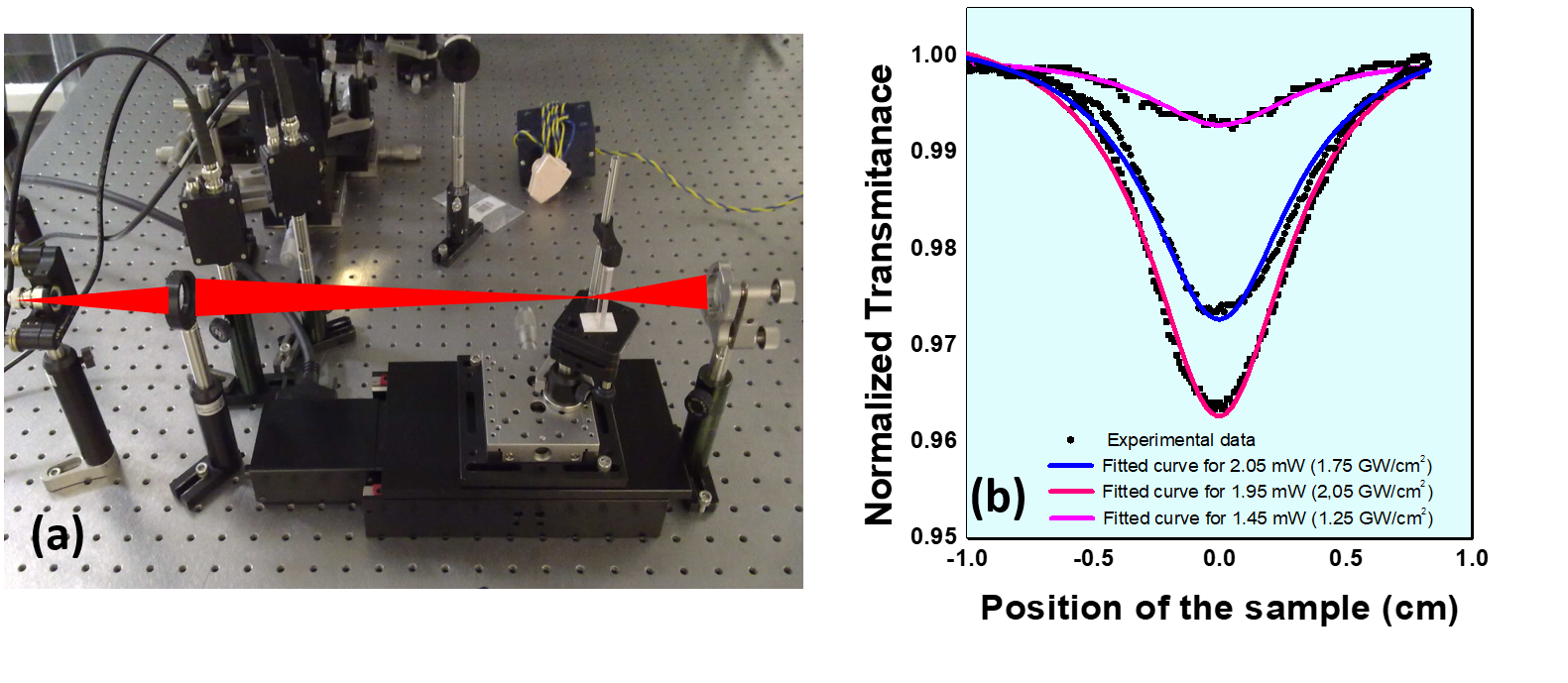}
\caption{Z-scan (a) open-aperture optical nonlinearity measurement setup and (b) acquired data for Bi$_2$Te$_3$ TISA}
\label{z-graph}
\end{figure}
\begin{figure}[h!]
\centering\includegraphics[width=15cm]{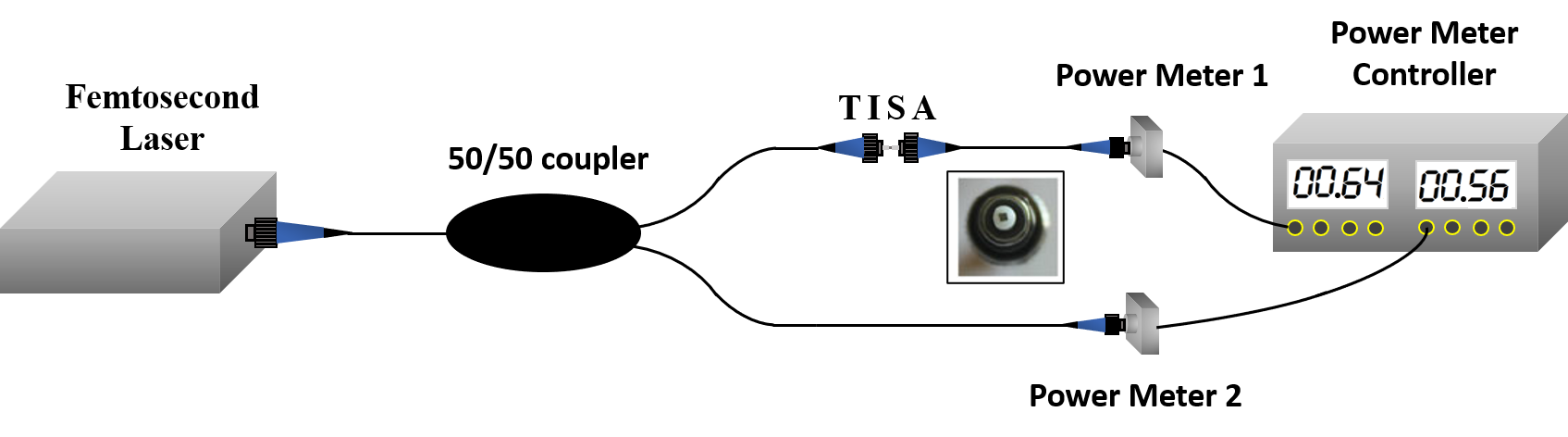}
\caption{Schematic diagram of optical nonlinearity measurement of TISA using balanced twin-detector technique}
\label{twin}
\end{figure}

\begin{figure}[h!]
\centering\includegraphics[width=10cm]{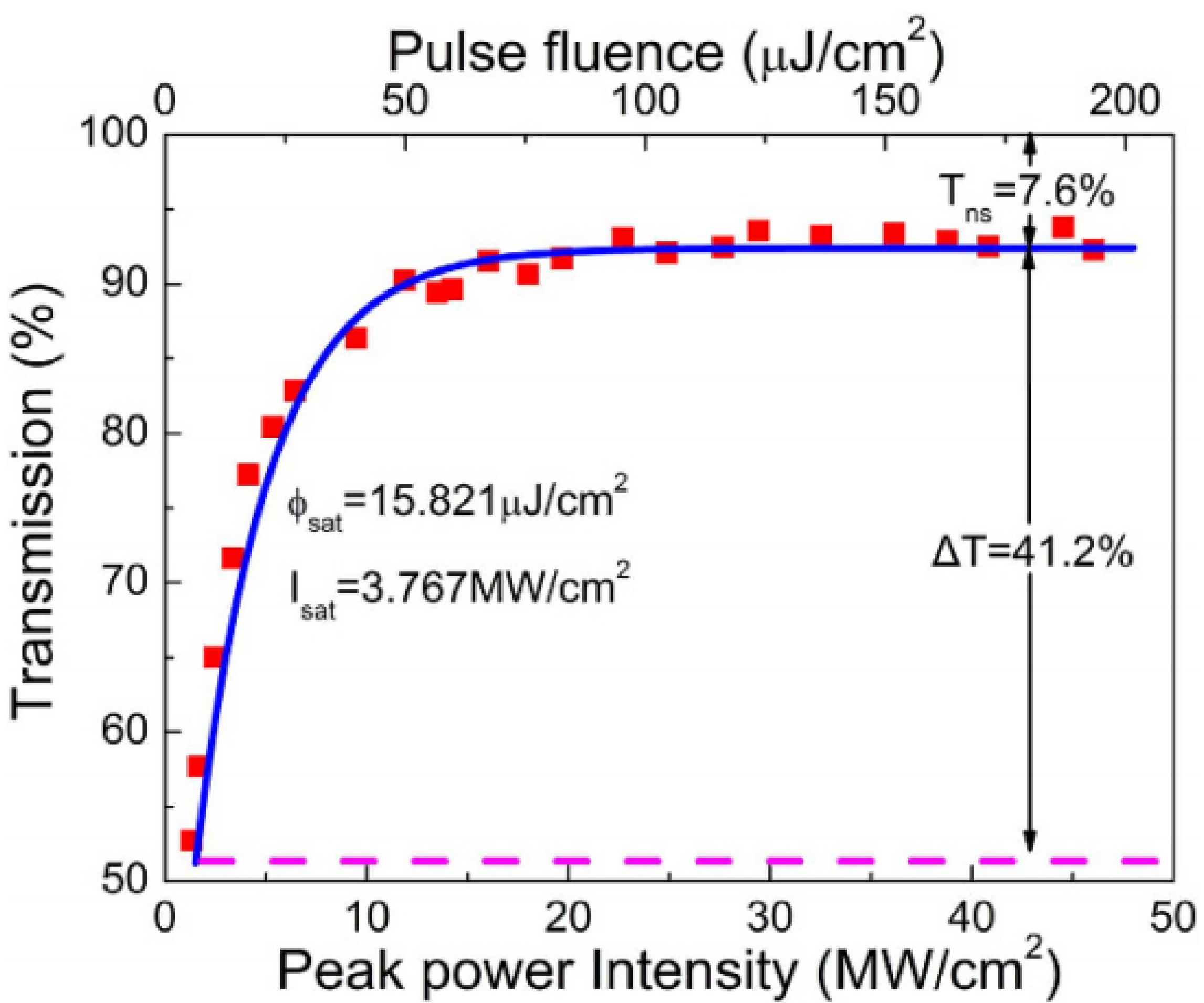}
\caption{Balanced twin detector technique nonlinear optical absorption profile. Reported with permission.\cite{li2016black} Fig. 2, 2016, Sci. Rep.}
\label{twin-profile}
\end{figure}

\subsection{Optical characterization of TISA}
The optical characterization of TISA ensures the saturable absorption properties, such as, modulation depth, non-saturable loss, saturation intensity and inverse saturable absorption. It is mainly a nonlinear optical measurement setup. Methods such as z-scan \cite{lu2013third} and balanced twin-detector measurement techniques \cite{wang2008wideband, chen2013large} are employed to find these properties.

In a typical z-scan setup (Fig. \ref{z-scan}), the material sample is placed in a linear translation stage followed by the detectors D$_1$ (for open aperture z-scan) and D$_2$ (for closed aperture z-scan). 
The sample is traversed along the light propagation direction (z-axis) where the incident ultrafast laser pulse varied dynamically as it comes through the lens and the z-axis fluence is calculated in terms of a function of the spot size. A reference signal (from D$_2$) is also taken to correct the measurement with respect to background. With the increase in the incident intensity, the saturable absorption by the material increases and it gradually saturates, which gives a typical saturable absorption curve (Fig. \ref{z-graph}(b)). 
The non-linear absorption coefficient of TI samples can be measured via open-aperture Z-scan technique (Fig. \ref{z-graph}(a)). To amend  effect of nonlinear absorption (e.g. two-photon absorption) over the nonlinear index measurement, the open-aperture method is generally used along with the closed-aperture z-scan method (see (Fig. \ref{z-scan}). 
The absorption coefficient($\alpha$) can be expressed as, $\alpha(I)=\alpha_0 +\beta I$, where, I is the laser intensity, $\alpha_0$ and $\beta$ is the linear non-linear absorption coefficient respectively. Z-scan curve can be defined by \cite{sheik1990sensitive},
\begin{equation}
T(z) = \sum\limits_{n = 0}^\infty 
\frac{{{{( - \beta I_{peak}^f{t_{eff}})}^n}}}
{(1 + {z^2}/z_0^2)^n}{(n + 1)^{3/2}}
\end{equation}

For n $\approx{1}$,

\begin{equation}
T(z) = - \frac{{\beta I_{peak}^f{t_{eff}}}}{{{2^{2/3}}(1 + {z^2}/z_0^2)}}\end{equation}
where, T(z) is the normalized transmittance, I$_{peak}^f$ is peak value of z-axis intensity at focal point, z is the sample position with respect to focal point, z$_0$ is the position of the focal point and the effective sample thickness is t$_{eff}$=(1- e$^{-\alpha_0t}$) / $\alpha_0$, t is  the actual thickness. The modulation depth ($\alpha_s$), maximum change in the transmission, can be derived from T(z) by the expression, $T(z)=1- \alpha _s /(1+ I/I_{sat})- \alpha_{ns}$, where, $\alpha_{ns}$ is the non-saturable loss, $I_{sat}$ is the saturable intensity and I is the incident intensity.

The balanced twin-detector techniques  \cite{wang2008wideband, chen2013large} is also a popular alternative power-dependent measurement technique to characterize the optical nonlinearity of SA. The schematic diagram of optical nonlinearity measurement of TISA using balanced twin-detector technique is shown in Fig. \ref{twin}.
A femtosecond optical pulse source can be used for this purpose. A 3 dB coupler is connected followed by the SA sample coupled in-between two connectors. The setup uses two separate power meters for comparison. Power achieved from the TISA is compared to the reference power. 
The variation of nonlinear transmission  w.r.t. to the peak intensity of a saturable absorber from the balanced twin-detector technique is shown in Fig.\ref{twin-profile}. It can be observed from this figure that the nonlinear transmission increases rapidly and finally saturates at a particular increased pulse fluence. Here, T$_{ns}$ and $\Delta$T are non-saturable loss and modulation depth, respectively.
However, balanced twin-detector method is not as accurate as the Z-scan technique but is easier to setup. 


\section{Modelocking with TIs}
Modelocking is one of the two techniques (the other one is Q-switching) to generate pulsed laser output from the fiber laser. It can successfully generate ultrashort pulses in picosecond and femtosecond orders.
Modelocking can be classified in accordance to the type of modelocker used, i.e., active or passive element. Active elements such as, acousto-optic devices, electro-optic modulator, Mach-Zehnder integrated-optic modulator, semiconductor electro-absorption modulator, etc. are used to modelock the laser.
Whereas, passive modelocker, also known as saturable absorber, can be further subdivided into artificial and real absorbers. 
Artificial types include cascaded second-order modelocking  \cite{mondal2016dynamic, mondal2014comparative}, nonlinear polarization evolution (NPE) \cite{fermann1993passive}, Kerr-lens modelocking \cite{cerullo1994self}, etc., while real domain includes SESAMs \cite{gomes2004picosecond}, 2D nanomaterials \cite{zhang2009large}, etc. 

Usually, a large number of longitudinal modes present inside the laser cavity depending on the wavelength and length of the cavity. But those modes are in the gain bandwidth spectrum, sustained in the cavity. These modes are spaced out in frequency domain and denoted by, $\Delta \nu = c/L_{opt}$, where, $L_{opt}$ is the total optical cavity length. When the modes are phase-locked, i.e., there is a constant phase difference between the neighboring modes, the modelocking happens. The ultrashort pulses can be generated due to modelocking when the cavity parameters such as linear/nonlinear loss, gain, dispersion and nonlinearity are adjusted properly. 

We can define modelocking mathematically by describing the master modelocking equation \cite{mukhopadhyay2013dual},
\begin{equation}
\frac{{\partial A}}{{\partial z}} + \frac{i}{2}{\beta _2}\frac{{{\partial ^2}A}}{{\partial {t^{_2}}}} - \frac{g}{2}\left( {A + \frac{1}{{\Omega _g^2}}\frac{{{\partial ^2}A}}{{\partial {t^{_2}}}}} \right) - \frac{\alpha }{2}A = i\left( {\gamma  + \frac{i}{2}l(t)} \right)|A{|^2}A
\end{equation}
where, A is pulse amplitude, $\alpha$ is loss parameter,$\beta_2$ is group velocity dispersion parameter, $\Omega_g$ is gain bandwidth, $\gamma$ is self-phase modulation parameter, g is the average saturated gain over the entire cavity, where,
$g = {g_0}/\left[ {1 + (2{I_{cir}}/{I_{sat}})} \right]$
where, 
${I_{cir}}\left( { = {\rm{ }}{{\left| {A\left( t \right)} \right|}^2}} \right)$
is the intensity circulating in the cavity,
${I_{sat}}\left( { = {\rm{ }}h{\omega _f}/{\sigma _L}} \right)$
is the saturation intensity. Here, $\sigma_L$ is the stimulated emission cross-section, $\omega_f$ is the operating frequency.

The small signal gain can be expressed as,
\begin{equation}
{g_0} = \frac{{2{\sigma _L}{\tau _L}{\eta _Q}{\eta _S}{\eta _{sys}}{P_{in}}}}{{\hbar \pi {\nu _p}{{\left( {MFD} \right)}^2}}}
\end{equation}
where, $\sigma_L$ is the stimulated emission cross section, $\tau_L$ is the fluorescence lifetime, $\eta_Q$ is the quantum efficiency, $\eta_S$ is the Stokes factor, $\eta_{sys}$ is the system efficiency, $P_{in}$ is the input pump power, $\hbar$ is the Planck’s constant, $v_p$ is the pump photon frequency and MFD is mode field diameter  \cite{mondal2016dynamic}. Here, $\it{l(t)}$ is the self-amplitude modulation parameter, i.e., intensity dependent loss which depends upon the intracavity peak intensity. This loss comes from saturable absorber which allows the central part of the modelocked pulse to be transmitted without loss but the pulse wings having low intensity experience loss. This time dependent loss can be expressed as,
\begin{equation}
l(t) = \frac{{{l_i}}}{{1 + \left[ {{{{{\left| {A(t)} \right|}^2}} \mathord{\left/
 {\vphantom {{{{\left| {A(t)} \right|}^2}} {{I_{sat}}}}} \right.
 \kern-\nulldelimiterspace} {{I_{sat}}}}} \right]}} \approx {l_i}\left[ {1 - \frac{{{{\left| {A(t)} \right|}^2}}}{{{I_{sat}}}}} \right]
\end{equation}

Passive modelocking is associated with saturable absorber since the early 1970s. 
Generally, the noise like spikes are generated above the noise floor due to the environmental disturbance during stimulated emission process in the resonator. 
Among all the spikes, the strongest one would get more gain from the gain medium and settle at the center of the gain bandwidth. 
When the spike passes through the gain medium, it produces in-phase longitudinal modes due to stimulated emission process and evolves as a modelocked pulse. 
At each round trip, that pulse, generated from the spike, roams around the resonator and experiences several effects like, gain, non-saturable loss, saturable loss, dispersion, nonlinearity, etc. in the cavity. It saturates the SA depending on its intensity.
The response time of the SA and the intra-cavity intensity will ultimately decide the ultrashort pulsewidth extracted from the laser resonator. 
When the spectral width of the laser pulse gets close to the gain bandwidth due to phase-locking of maximum number of modes, the shortest pulse, i.e., bandwidth limited pulse are achieved. 

A good saturable absorber having short excited state lifetime is required to generate ultrashort pulses. 
With the advancement of nanotechnology, 2D materials have been used as SAs since the discovery of the nonlinear optical properties of graphene.
This motivates researchers to continue their search for better 2D materials which could provide ultrashort pulses in the entire infrared regime.
Topological insulators provide the Dirac cone similar to that of graphene. 
It led to TIs being introduced in the field of ultrafast pulse generation. Generally, TIs have comparatively flat transmission curve from near-infrared to mid-infrared wavelength region, hence it is suitable as a SA for Yb, Er and Tm-doped fiber lasers by adjusting the saturable absorption parameters.
François Bernard and Han Zhang \cite{bernard2012towards} investigated the nonlinear optical properties of Bi$_2$Te$_3$ for the first time and identified that these types of two-dimensional materials can be used for several photonic applications. 
As discussed earlier, various TIs have been used as ultrafast saturable absorber. 
Among them, Bi$_2$Te$_3$, Bi$_2$Se$_3$, Sb$_2$Te$_3$, bismuthine, tellurene and AuTe$_2$Su$_{4/3}$ are the promising candidate for ultrafast SAs.
To get the pure and stable modelocking, one needs high modulation depth, low saturation intensity and high signal to noise ratio.
Topological insulator based saturable absorbers have role in switchable as well as in multi-wavelength operation in modelocked fiber lasers. 
In comparison with other 2D materials, topological insulator has unique property of low saturation intensity and broad effective bandwidth which support dual, multi-wavelength and harmonic modelocking. 

\begin{figure}[h!]
\centering\includegraphics[width=15cm]{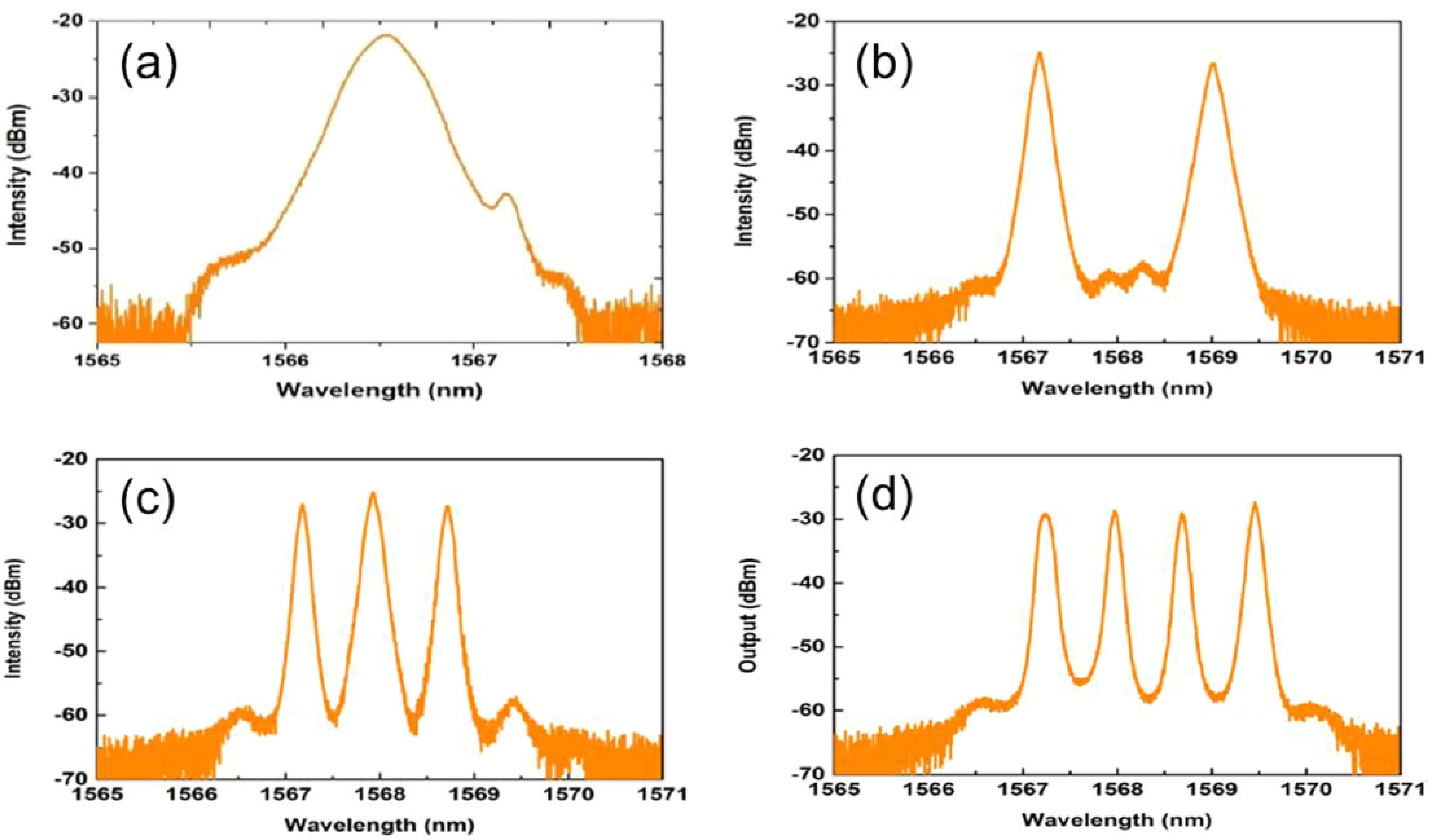}
\caption{Optical spectrum of multi-wavelength modelocking regime for (a) fundamental; (b) dual-wavelength; (c) triple-wavelength; (d) four-wavelength modelocking . Reported with permission (Fig. 4(a)-(d), \cite{guo2015topological}) 2015, JAP.}
\label{fig:guo2015topological(1)}
\end{figure}

\begin{figure}[h!]
\centering\includegraphics[width=15cm]{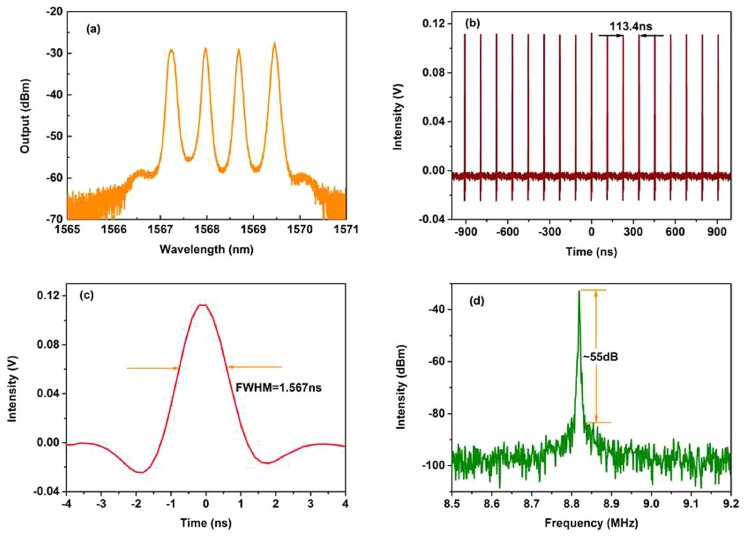}
\caption{Optical characteristics at four-wavelength mode-locking regime. (a) output optical spectrum; (b) pulse train; (c) zoomed in single-pulse profile; (d) RF spectrum. Reported with permission (Fig. 5(a)-(d), \cite{guo2015topological}) 2015, JAP.}
\label{fig:guo2015topological}
\end{figure}

The origin of dual wavelength mode-locking in TISA based fiber laser lies in the amount of birefringence produced by the polarization controller. 
The polarization controller induced birefringence provides the phase shift to the orthogonal polarizations.
As the phase shift between the two orthogonal polarizations becomes large, the electric field associated with those polarizations are reaching to the gain medium at different times.
The frequencies which overcomes the linear cavity losses, consisting of both the polarizations, occupy different positions in the gain medium. 
Due to continuous pumping to the gain medium and its faster recovery time than the repetition rate, the gain medium is always in inverted condition.
As the relaxation time of the TISA is faster than the separation of the two bunch of pulses with two separate polarizations and wavelengths, generates different sets of mode-locked pulses.
In Fig.\ref{fig:guo2015topological(1)}, the optical spectrum of fundamental, dual, triple and four-wavelength modelocking has been represented \cite{guo2015topological}.The corresponding pulse train, single-pulse profile and RF spectrum has been depicted in Fig.\ref{fig:guo2015topological}.
The theoretical investigations and experimental evidences having dual and multi-wavelength operation of TISA have been discussed in the recent literature  \cite{guo2015topological, liu2014dual, yasim2020switchable, lee2020numerical, song2019dual}. 

High repetition rate pulse fiber lasers are considered as light sources in the field of optical fiber communication systems, astronomical frequency combs, laser spectroscopy, etc. 
To increase the repetition rate without reducing the cavity length requires harmonic mode-locking state. Many saturable absorbers like CNT, Graphene and non-linear polarization evolution (NPE)  \cite{zhu2013passive,zhang2013sub,grudinin1997passive,sobon201110}, have been demonstrated for harmonic mode-locking of fiber lasers with high repetition rate. Z.C. Luo et al.  \cite{luo20132} were the first to test the HML with TISA. 
Compared to other 2D materials, a TI has insulating properties in the interior and conducting states on its surface which makes it suitable for broadband saturable absorption. 
To generate HML laser pulses, high nonlinearity is a must which is connected to the intensity dependent refractive index of a material. 
Before Z. C. Luo et al., people tested TISAs over quartz plate \cite{zhao2012wavelength, yu2013topological} or at the fiber end-facet  \cite{zhao2012ultra} which do not provide much saturable absorption due to short interaction length between the light and TISA. 
However, tapered fiber and micro-structured fiber TISA schemes evidently showed high harmonic modelocking. 
This indicates that such types of TISA can generate high nonlinearity with excellent saturable absorption capability for ultrafast fiber lasers.

\begin{figure}[h!]
\centering\includegraphics[width=15cm]{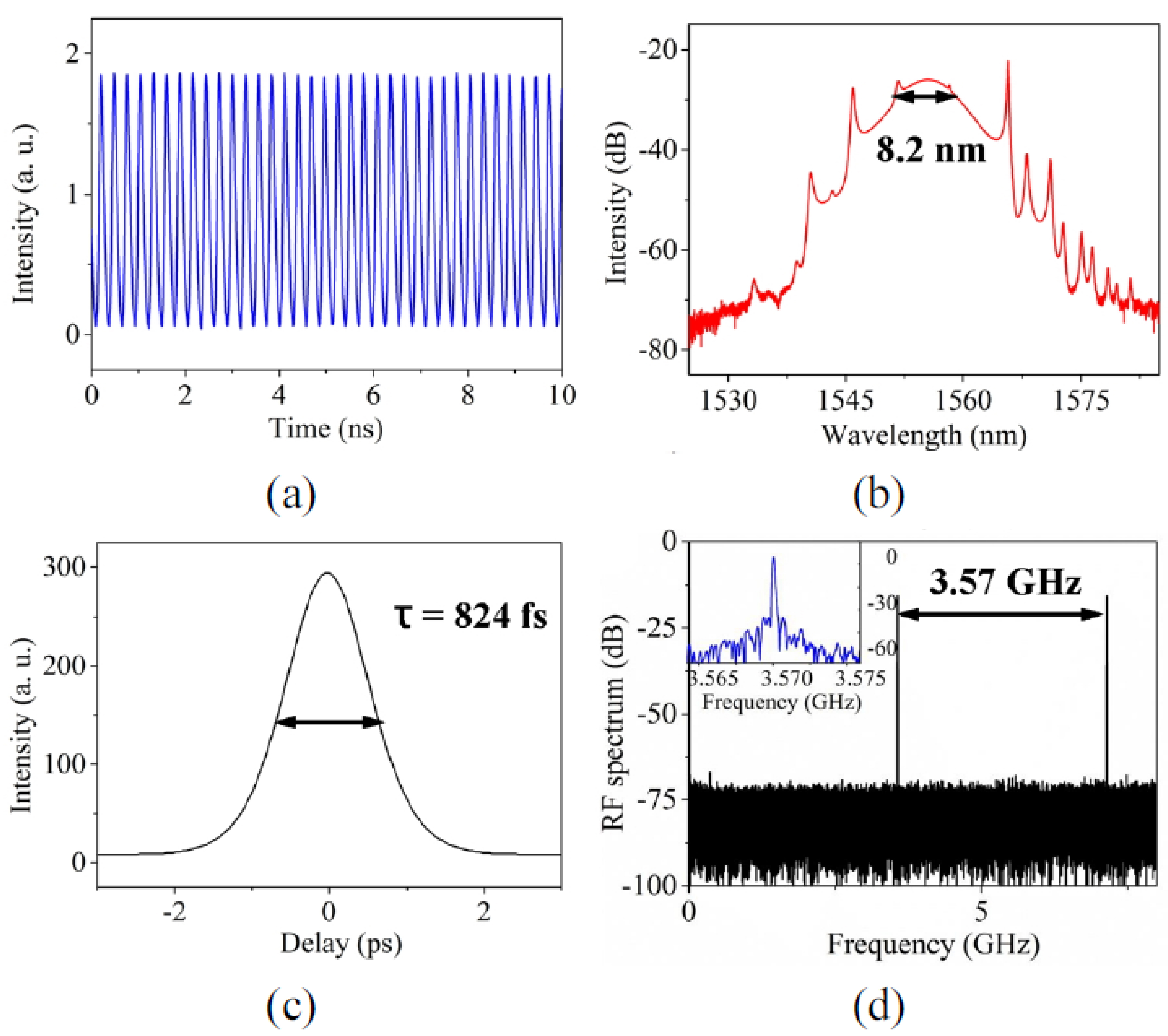}
\caption{Optical characteristics of harmonic modelocking regime. (a) pulse train; (b) optical spectrum; (c) autocorrelation trace; (d) RF spectrum. Reported with permission (Fig. 5(b), \cite{li2016vector}) 2016, OSK.}
\label{fig:li2016vector}
\end{figure}

According to the soliton area theorem  \cite{agrawal2000nonlinear}, as the pump power increases, the limiting factor of peak power generates the multiple quantized soliton pulses per round trip. By careful adjustment of pump power and meticulous alignment of the polarization controller, the pulses moving inside the laser cavity experience equalized peak power, pulse interval and high non-linear effect. As a result, stable harmonic modelocked pulses can be generated in N$^{th}$ harmonic. 

The optical spectrum characteristics of the highest stable harmonic achieved till date \cite{li2016vector} has been represented in Fig.\ref{fig:li2016vector}, where the pulse train, optical spectrum, autocorrelation trace and RF spectrum has been shown.
From the results of various researchers, it has been analyzed how topological insulators like Bi$_2$Se$_3$, Bi$_2$Te$_3$ and Sb$_2$Te$_3$ played a vital role in the dual wavelength mode-locking and also harmonic mode-locking in fiber lasers. The outcome of these dual wavelength and harmonic mode-locking has applications in the field of spectroscopy, bio medicine, sensing research, optical fiber communication systems, astronomical frequency combs, laser spectroscopy and etc.

Significant optical parameters along with fabrication and integration techniques for all the TISA-based modelocked fiber lasers (both fundamental and HML) have been tabulated in Tables \ref{table2} and \ref{table3} respectively. The direct comparison among different results shown in these tables makes it easier to depict which TISA material and fabrication process provides the best modelocked laser in terms of output power, wavelength, modelocking threshold, pulse duration, etc. 

The best modelocking parameters for TISA-based fiber lasers have been tabulated in Table \ref{table4}. From this table we have found that the shortest pulse duration acquired from Sb$_2$Te$_3$ material is $\sim$70 fs \cite{liu201670} as the nonlinear polarization evolution technique (NPE) as well as tapered-fiber TISA have been used as saturable absorbers. It means, in the above mentioned work, the advantage of both imaginary and real SA have been utilized to achieve shortest pulse duration.
The intra-cavity pulse duration mainly depends on the bandwidth of SA. The bandwidth of SA mostly depends on the available DOS and the relaxation time at that particular wavelength range. Moreover, the gain bandwidth is also limited by the amplified spontaneous and stimulated emissions. It is found in the literature that the gain bandwidth of erbium, ytterbium and thulium doped gain fibers are $\sim$ 40 nm \cite{wysocki1997broad}, $\sim$55 nm \cite{paschotta1997ytterbium} and $\sim$100 nm\cite{yan2016widely} respectively. From our observation supported by the literature review, among all the fabrication methods, PLD provides shorter pulsewidth due to better control of growth of atomic layers. However, with the same deposition technique (i.e. PLD) having same TI material (i.e. Bi$_2$Te$_3$), the achievable pulsewidth from the fiber laser cavity differs far from each other at different operational wavelengths. For example, in reference \cite{lu2015yb} with Yb-doped fiber laser, they achieved pulsewidth of 317 ps at center wavelength of 1052.5 nm, whereas, 452 fs has been generated at 1558.5 nm \cite{chen2016high} with same PLD based Bi$_2$Te$_3$ TISA. It may be noticed that the gain bandwidth of Yb-doped fiber \cite{paschotta1997ytterbium} is a bit higher than Er-doped fiber \cite{wysocki1997broad}. Therefore, a conclusion can be drawn that the pulsewidth is restricted by the inherent property of SA (i.e. DOS and relaxation time) at two different wavelength ranges.

\newpage
\begin{sidewaystable}
\caption{Performance summary of optical parameters for TI based saturable absorber in fundamental modelocking:}
\begin{tabular}{|p{1.3cm}|p{1.8cm}|p{1.6cm}|p{1.8cm}|p{1.0cm}|p{1.5cm}|p{1.6cm}|p{1.3cm}|p{1.5cm}|p{1.4cm}|p{0.95cm}|p{1.2cm}|p{1.2cm}|p{1cm}|}
\hline
Wave- length (nm) & Topological Insulator & Fabrica- tion Technique & Threshold pump power & Pulse width & Saturation intensity /power /fluence & Modulation depth & Non-Saturable loss & Repetition rate & Maximum Output power & SNR & 3-dB Spectral bandwidth & Time bandwidth product & *Ref. \\
 &  & & (mW) & (ps) & (MW/cm$^2$) & (\%) & (\%) & (MHz) & (mW) & (dB) & (nm) &  &  \\
 \hline

1558.4            & Bi$_2$Te$_3$    & HTE      & 70                        & 1.21                                                             & 480                                                                           & $\sim$95.3                                                            & ---                                                                      & 1.21                                                                  & ---                                                                        & ---                                                       & 2.69                                                                    & 0.40                   & \cite{zhao2012ultra}                                                   
                                      \\
1570            & p-type Bi$_2$Te$_3$ & MT  &---                & 0.403                                                            & 0.058                                                                         & 3                                                                    & $\sim$10                                                                & 28.5                                                                  & ---                                                                        & ---                                                       & 6.86                                                                    & 0.315    
                                
& \cite{lin2014soliton}         \\ 
1935            & Bi$_2$Te$_3$    & ME      & ---                        & 0.795                                                            & ---                                                                            & $\sim$20.6                                                            & $\sim$17.5                                                              & $\sim$27.9                                                            & $\sim$1                                                                  & 76                                                       & $\sim$5.64                                                              & 0.35                  & \cite{jung2014femtosecond}                   \\    
1558.12         & Bi$_2$Te$_3$    & HTE      & 401                       & 1.28                                                             & ---                                                                            & ---                                                                    & ---                                                                      & 1.216                                                                 & 0.59                                                                      & ---                                                       & 9.3                                                                     & 0.325                  & \cite{chen2014formation}                                        \\
1562            & Bi$_2$Te$_3$    & Polyol      & 35                        & 0.82                                                             & $\sim$10                                                                     & $\sim$2.4                                                             & $\sim$65.7                                                              & 12.5                                                                  & ---                                                                        & 58                                                       & 3.3                                                                     & $\sim$0.332            & \cite{luo2014observation}                                        \\
1057.82         & Bi$_2$Te$_3$     & ME     & 200                       & $\sim$230                                                        & 92 W                                                                            & $\sim$1.8                                                             & $\sim$85                                                                & 1.44                                                                  & 0.86                                                                      & $\sim$77                                                 & 3.69                                                                    & 227.5369          & \cite{chi2014all}                                        \\
1052.5          & Bi$_2$Te$_3$      & PLD    & 230                       & 317                                                              & ---                                                                            & 10                                                                    & 76                                                                      & 19.8                                                                  & 3                                                                         & ---                                                       & 1.245                                                                   & 106.8823             & \cite{lu2015yb} 
\\
1063.4          & Bi$_2$Te$_3$    & PLD      & 220                       & 5470                                                             & ---                                                                            & 10                                                                    & ---                                                                      & $\sim$6.16                                                                  & 12.4                                                                      & 60                                                       & 2.24                                                                    & 3250                  & \cite{li2017high}                                        \\
1558.418        & Bi$_2$Te$_3$     & CVD     & 115                       & 3220                                                             & 5.8                                                                           & 10.8                                                                  & 86.5                                                                    & 1.704                                                                 & 40.37                                                                     & \textgreater{}60                                         & 1.696                                                                   & 674.58                 & \cite{wei2019large}                                       \\
1558.5          & Bi$_2$Te$_3$      & PLD    & 108.3$\mu$J/cm$^2$                       & 0.4523                                                            & 19                                                                            & 6.5                                                                   & 38.4                                                                    & 20.1                                                                  & 1.82                                                                      & ---                                                       & 6.72                                                                    & 0.373                  & \cite{chen2016high}                                        \\
$\sim$1560      & Bi$_2$Te$_3$     & ME     & 88                        & 2700 – 12800                                                     & 44-55 W                                                                            & $\sim$15.7-9.9                                                            & 17                                                                      & 1.7                                                                   & 6.8–32.9                                                                & $\sim$51                                                 & $\sim$5.6                                                               & 1863.9-8836.3       & \cite{lee2016pulse}                                        \\
1557-1565       & Bi$_2$Se$_3$ & Polyol         & 65                        & 1.57                                                             & 490                                                                           & \textless 98                                                          & $\sim$0.8                                                               & 1.21                                                                  & ---                                                                        & ---                                                       & 1.79                                                                    & $\sim$0.35             & \cite{zhao2012wavelength}  
            \\
1031.7          & Bi$_2$Se$_3$     & Polyol     & 153                       & 46                                                               & 580                                                                           & $\sim$5.2                                                             & $\sim$56.5                                                              & 44.6                                                                  & 33.7                                                                      & 58                                                       & 2.5                                                                     & 32.4125                 & \cite{dou2014mode}                                        \\
1557.5          & Bi$_2$Se$_3$    & Polyol    & 25                        & $\sim$0.66                                                       & 12                                                                            & $\sim$3.9                                                             & $\sim$67.5                                                              & 12.5                                                                  & 1.8                                                                       & 55                                                       & 4.3                                                                     & 0.351                  & \cite{liu2014femtosecond}                                       \\

1600            & Bi$_2$Se$_3$    & Polyol      & 130                       & $\sim$0.36                                                       & 12$\mu$J/cm$^2$                                                                            & 5.2                                                                   & $\sim$38                                                                & 35.45                                                                 & 0.86                                                                      & \textgreater{}56                                         & 7.9                                                                     & 0.333                  & \cite{li2015band}                                        \\
 1554.56          & Bi$_2$Se$_3$    & OD       & 25                        & 0.908                                                            & 15 mW                                                                            & 5                                                                     & $\sim$66                                                                & 20.27                                                                 & 5.5                                                                       & 80                                                       & 7.91                                                                    & 0.607                  &   \cite{gao2015stable}                             \\
   1571            & Bi$_2$Se$_3$     &  QST     & 60                        & 0.579                                                            & 3200                                                                          & 88                                                                    & ---                                                                      & 12.54                                                                 & 1.59                                                              & 54                                                       & 5                                                                       & 0.352                  & \cite{xu2017bilayer}                                       \\
1912.12         & Bi$_2$Se$_3$     & ME     & $\sim$200                 & $\sim$0.835                                                      & 28.6W                                                                            & $\sim$13.4                                                            & $\sim$52.5                                                              & $\sim$18.37                                                           & ---                                                                        & $\sim$65                                                 & $\sim$4.87                                                              & $\sim$0.337            & \cite{lee2018femtosecond}                                       \\
1559            & Bi$_2$Se$_3$     & OD      & 60                        & 7.564                                                            & ---                                                                            & 5                                                                     & $\sim$66                                                                & 7.04                                                                  & $\sim$75                                                                  & 70                                                       & 26                                                                      & ---                  & \cite{gao2015stable} \\
1558.6          & Sb$_2$Te$_3$   & ME       & 45                        & 1.8                                                              & ---                                                                            & ---                                                                    & ---                                                                      & 4.75                                                                  & 0.5                                                                       & \textgreater{}60                                         & 1.8                                                                     & 0.405                  & \cite{sotor2014mode}                                         \\
1565            & Sb$_2$Te$_3$    &ME      & 30                        & 0.128                                                            & 31                                                                            & 6                                                                     & 43                                                                      & 22.32                                                                 & 1                                                                         & 65                                                       & 30                                                                      & 0.47                   & \cite{sotor2014sub}                                     
           \\

1556            & Sb$_2$Te$_3$    & LPE      & 44                        & 0.449                                                            & 106                                                                           & 3.9                                                                   & 87                                                                      & 22.13                                                                 & 0.9                                                                       & 74                                                       & 6                                                                       & 0.333                  & \cite{boguslawski2014mode}                                        \\

1558            & Sb$_2$Te$_3$     & PMS     & 90                        & 0.167                                                            & 2600                                                                          & 5.3                                                                   & $\sim$17.5                                                              & 25.38                                                                 & 5.34                                                                      & 68                                                       & 34                                                                      & 0.7                    & \cite{boguslawski2015dissipative}                                        \\
 
1542            & Sb$_2$Te$_3$     & PLD     & 91                        & 0.070                                                            & 175                                                                           & 7.42                                                                  & $\sim$29.2                                                              & 95.4                                                                  & 63                                                                        & 65                                                       & 63                                                                      & 0.5564                 & \cite{liu201670}                                       \\

1561            & Bismuthene   & SE    & 50                        & 0.193                                                            & 48.2                                                                          & 5.6                                                                   & 62.3                                                                    & 8.85                                                                  & 5.6                                                                       & $\sim$55                                                 & 14.4                                                                    & $\sim$0.342            & \cite{guo2018sub}                                        \\
1558.8          & Tellurene   & LPE    & 80                        & 1.03                                                             & 1.06                                                                          & 35.64                                                                 & 25.91                                                                   & 3.327                                                                 & 3.69                                                                      & $\sim$37.95                                              & 2.91                                                                    & 0.37                   & \cite{liu2020ultrathin}                                        \\
1557.53         & AuTe$_2$Se$_\frac{4}{3}$   & SF   & 93                        & 0.1477                                                           & ---                                                                            & 65.58                                                                 & 18.13                                                                   & 69.9                                                                  & 21.4                                                                      & 91                                                       & 30.26                                                                   & 0.5523                 & \cite{liu2020ultrafast}   \\

\hline
\end{tabular}
\label{table2}
\end{sidewaystable}

\clearpage

\newpage
\begin{sidewaystable}
\caption{Performance summary of optical parameters for TI based saturable absorber in dual-wavelength and harmonically modelocked fiber laser:}
\begin{tabular}{|p{1.4cm}|p{1.6cm}| p{1.6cm}|p{1.4cm}|p{1.2cm}|p{1.5cm}|p{1.6cm}|p{1.3cm}|p{1.5cm}|p{1.4cm}|p{0.95cm}|p{1.2cm}|p{1.3cm}|p{1.0cm}|}
\hline
Wave- length (nm) & Topological Insulator & Fabrica- tion Technique & Threshold pump power & Pulse width & Saturation intensity /power /fluence  & Modulation depth & Non-Saturable loss & Repetition rate & Maximum Output power & SNR & Spectral bandwidth & Time bandwidth product & *Ref. \\
 &  &  & (mW) & (ps) & (MW/cm$^2$) & (\%) & (\%) & (MHz) & (mW) & (dB) & (nm) &  &  \\
 \hline

1557.4 \& 1559.4                                                  & Bi$_2$Te$_3$   & HTE    & 53 \& 116                                                                    & 4.09 \&   3.17                                                   & 0.02 mW                                                                              & $\sim$16.3                                                            & $\sim$38.1                                                                & 5.1(F), 388(H) \&   239(H)            & ---                                                                          & ---               & 1.12 \&   1.75                                                            & 0.566 \&   0.6843        & \cite{liu2014dual}   \\
1558.5                                                          & Bi$_2$Te$_3$    & HTE   & 7.5 (F), 26.5 (H)                                                                 & 1.22(F), 2.49(H)                                                              & ---                                                                              & $\sim$1.7                                                             & $\sim$69.9                                                                & 4.88 (F) 2040 (H)                & 5.02                                                                         & ---               & 0.95(F), 1.08(H)                                                                      & 0.332                     & \cite{luo20132}    \\
1555.9                                                          & Bi$_2$Te$_3$     & ME  & 18(F), 22(H)                                                                    & $\sim$0.63 –0.7(H)                                            & $\sim$ 61 W                                                                              & $\sim$3.75                                                            & $\sim$45.2                                                                & 14.07(F) –   773.85(H)        & $\sim$1.3                                                                   & 46.3(F) – 63(H)        & $\sim$4.5                                                                 & $\sim$0.35               & \cite{lee2015femtosecond}   \\
1909.5                                                          & Bi$_2$Te$_3$    & OD   & $\sim$315                                                                   & 1.26                                                             & ---                                                                              & 9.8                                                                   & $\sim$53.5                                                                & 21.5                    & $\sim$60                                                                    & \textgreater{}50(F) & 3.43(H)                                                                      & 0.356                    & \cite{yin2015soliton}  \\
1562.4(F), 1564(H) \& 1564.1(H)                                                            & Bi$_2$Te$_3$     & PLD  & $\sim$35                                                                    & 0.32(F), 0.609(H) \& 0.92(H)                                                             & 28                                                                              & 6.2                                                                   & 20                                                                        & 17.34(F), 2230(H) \& 2950(H)                    & 0.56(F), 43.5(H) \& 45.3(H)                                                                        & ---               & 8.2(F), 4.55(H) \& 3.3(H)                                                                      & 0.323(F), 0.34(H) \& 0.37(H)                    & \cite{yan20152} \\
1560.88                                                         & Bi$_2$Te$_3$    & HTE+PLD   & 28                                                                          & 1.754                                                            & ---                                                                              & 31                                                                    & $\sim$68.6                                                                & 15.6 \& 3125                    & 0.5(F) \& 6.4(H)                                                                         & ---               & 2.12(F) \& 1.52(H)                                                              & 0.328                    & \cite{jin20183}\\
1561.6 \& 1562.1                                                  & Bi$_2$Se$_3$    & LPE   & 80                                                                          & 13620 – 25160                                                      & 26                                                                              & 4.1                                                                   & $\sim$69                                                                  & 3.54                    & 2.1-10                                                                   & $\sim$80         & 0.25                                                                      & 418.8 \&   773.3         & \cite{guo2015dual} \\
1567.5 \& 1570                                                  & Bi$_2$Se$_3$     & LPE  & 85.5                                                                        & $\sim$22                                                         & $\sim$25                                                                        & 3.8                                                                   & ---                                                                        & 8.83                    & 9.7                                                                         & $\sim$55         & $\sim$0.8 \& $\sim$0.12                                                   & $\sim$0.322              & \cite{guo2015topological} \\
1566 \& 1567   & Bi$_2$Se$_3$   & LPE     & 97.3                                                                        & ---                                                           & $\sim$25                                                                        & $\sim$3.8                                                             & $\sim$68                                                                  & 1.086                   & ---                                                                          & $\sim$62         & ---                                                                        & ---                       & \cite{guo2015observation}  \\
1610 & Bi$_2$Se$_3$ & HTE & $\sim$1200 & $\sim$3.1(F) \& $\sim$2.76(H) & --- & $\sim$4.3 & $\sim$60.7 & 7.04(F) \& 640.9(H) & 308 & --- & 0.91(F) \& 1.06(H) & $\sim$0.32 & \cite{meng2015high} \\

1532 \& 1557                                                      & Bi$_2$Se$_3$    & Polyol   & 40 \& 120                                                                   & 1.7 \& 0.5                                                      & 77                                                                              & 12                                                                    & 48                                                                        & 38.718 \&   38.716      & ---                                                                          & \textgreater{}50 & 1.5 \& 6.5                                                                & 0.325 \& 0.402           & \cite{li2017analysis} \\
1550                                                            & Bi$_2$Se$_3$    &  OD  & 18                                                                          & 0.958(F) \& 0.824(H)                                                            & ---                                                                              & 2.3                                                                   & $\sim$56.2                                                                & 10.03 \& 3570                    & 11.5                                                                        & 70               & 7.3(F) \& 8.2(H)                                                                       & 0.571                    & \cite{li2016vector} \\

1558                                                            & Sb$_2$Te$_3$    & ME   & 35                                                                          & 1.9(F), 2.2(H)                                                       & ---                                                                              & ---                                                                    & ---                                                                        & 3.75(F), 304(H)                     & 0.5(F), 4.5(H)                                                                         & 65(F), 55(H)               & 1.58                                                                      & 0.42                     & \cite{sotor2014harmonically} \\

1562.71                                                         & Sb$_2$Te$_3$     & PMS  & 108 – 520                                                                     & 1.61                                                             & ---                                                                              & 1.5                                                                   & 64.9                                                                      & 13.2(F), 105.6(H) \&  607.2(H)           & 5.09(H) \& 25.54(H)                                                                        & 77               & 2.31                                                                      & 0.4568                   &\cite{wang2019generation}\\
\hline
\end{tabular}
\label{table3}
\end{sidewaystable}
\clearpage

\begin{table}
\caption{Best performance summary of optical parameters for TISA in modelocked fiber lasers:}
\centering
\arrayrulecolor{black}
\begin{tabular}{|p{2cm}|p{1.4cm}| p{1.6cm}|p{1.7cm}|p{1.6cm}|p{1.95cm}|p{1.4cm}|p{1.8cm}|}
\hline
Parameters                                   & Values       & TISA \ material             & Wavelength
  (nm)        & Fabrication
  process        & Integration
  technique             & Ref.                                                             & Remarks                                              \\ 
\hline
Minimum Pulsewidth (ps)                      & 0.07         & \multirow{3}{*}{Sb$_2$Te$_3$
  ~} & \multirow{3}{*}{1542}    & \multirow{3}{*}{PLD
  ~
  ~} & \multirow{3}{*}{Tapered fiber}      & \multirow{3}{*}{
      \cite{liu201670}}          & \multirow{3}{*}{NPE + TISA, hybrid$^1$}               \\ 
\cline{1-2}

Highest   Repetition rate (MHz)                & 95.4 (F)
  ~ &                             &                          &                              &                                     &                                                                  &                                                      \\ 
\hline
Highest Repetition rate - HML (MHz)         & 3125 (H)     & \multirow{2}{*}{Bi$_2$Te$_3$} & \multirow{2}{*}{1560.88} & \multirow{2}{*}{HTE+PLD}     & \multirow{2}{*}{Tapered fiber}  & \multirow{2}{*}{  \cite{jin20183} }        & \multirow{2}{*}{Microfiber and high nonlinearity}  \\ 
\cline{1-2}
Highest Harmonic                             & 200$^{th}$        &                             &                          &                              &                                     &                                                                  &                                                      \\ 
\hline
Lowest Repetition rate (MHz)
  ~             & 1.21         & \multirow{3}{*}{Bi$_2$Se$_3$}     & \multirow{3}{*}{1564.6}  & \multirow{3}{*}{Polyol}      & \multirow{3}{*}{Quartz 
   substrate} & \multirow{3}{*}{
  \cite{zhao2012wavelength}} & Cavity length
  = 247.93 m                           \\ 
\cline{1-2}\cline{8-8}
Minimum Non-Saturable loss (dB)              & 0.8          &                             &                          &                              &                                     &                                                                  & High
  saturable loss                                \\ 
\cline{1-2}\cline{8-8}
Maximum Modulation depth (\%)                & 98           &                             &                          &                              &                                     &                                                                  & Strong
  saturable absorption, self-starting         \\ 
\hline
Maximum Output
  Power (mW)                  & 308           & Bi$_2$Se$_3$                            & 1610                         & HTE                             & Tapered fiber                                    & \cite{meng2015high}                                                                 & High nonlinearity and high pump power                                                     \\ 
\hline
Minimum Saturation intensity /power /fluence & 1.06         & Tellurene                   & 1558.8                   & LPE                          & Tapered fiber                       &  \cite{liu2020ultrathin}                        & Ultra-thin,
  less energy levels, fast saturation    \\ 
\hline
Minimum Modulation depth (\%)                & 1.5          & Sb$_2$Te$_3$                      & 1562.71                  & PMS                          & Tapered fiber                       &  \cite{wang2019generation}                            & Quick saturation,
  leads to multi-pulse solitons    \\ 
\hline
Time-bandwidth
  Product (sech$^2$)             & 0.323        & Bi$_2$Te$_3$                      & 1562.4                   & PLD                          & Tapered fiber                       &  \cite{yan20152}                       & Optimum
  intra-cavity dispersion and nonlinearity   \\
\hline
\end{tabular}
\arrayrulecolor{black}
\label{table4}
\\
. 

$^1$hybrid refers to the state of cavity configuration, light propagates in fibers as well as in free-space in laser cavity.
\end{table}
 
\clearpage

Even though the gain bandwidth of thulium-doped fiber is higher than that of Yb and Er-doped fibers, TISA fabricated by PLD  has not been tested at $\sim$ 2 $\mu$m wavelength in the literature. 
Moreover, it can be seen from Table \ref{table2} that Bi$_2$Se$_3$, fabricated by polyol method, could generate pulses as short as $\sim$360 fs. However, PLD-based Bi$_2$Se$_3$ TISA has still not been used in fiber lasers at any of the wavelength ranges.
Also, minimum modulation depth \cite{wang2019generation} and highest repetition rate (among the reported fundamental modelocking works) \cite{liu201670} are found to be better in Sb$_2$Te$_3$ than in Bi$_2$Te$_3$ and Bi$_2$Se$_3$. 
It is also found that the maximum output power having HML is $\sim$308 mW at 1610 nm with Bi$_2$Se$_3$ based modelocked EDFL \cite{meng2015high}. They used tapered fiber based TISA in this case with very high pump power of 5 watts.  
Furthermore, in Ref. \cite{liu201670}, they realized modelocking in a hybrid ring cavity configuration, hence optical path is less and it is reflected through the high repetition rate.
However, for highest repetition rate (3.125 GHz) and maximum harmonic order (200$^{th}$) among all HML papers has been reported by  L. Jin et al. \cite{jin20183}. Here, they have used microfiber based TISA, therefore, the intra-cavity nonlinearity is very high; due to this, highest harmonic modelocking was found.
Another interesting observation from the table \ref{table4} is that the lowest repetition rate (1.21 MHz), minimum non-saturable loss (0.8 dB) and highest modulation depth (98\%) has been reported while using Bi$_2$Se$_3$-based TISA \cite{zhao2012wavelength}. The long cavity length of the entire fiber setup (247.93 m) resulted in such low repetition rate. 

The optical properties of TISA, such as modulation depth, saturation intensity, recovery time, non-saturable loss and damage threshold have a direct connection to the stability of laser pulses. Spectrum bandwidth is proportional to modulation depth, whereas pulsewidth is inversely proportional to it. Hence, it can be inferred that, for stronger saturable absorption higher modulation depth is required. On the other hand,  group velocity dispersion (GVD), third order non-linearity and saturation gain of an active fiber laser cavity require low modulation depth. The nonlinear transmission behaviour is a function of intracavity intensity, measured by z-scan technique, can be explained by the equations (1) and (2) defined earlier. It has been observed that TISA is an intrinsic high modulation depth SA and it inherently suppresses the chance of pulse-breaking phenomena. Hence, it supports high energy dissipative solitons. 
TISA having lower saturation intensity, saturates faster, allows more number of cavity modes which extract gain during the cavity round trips. 
During the round trips, when the amplitude of the laser pulse gets close to the saturation due to continuous pumping, stable modelocked pulse is achieved. After a certain intracavity power, pulses cannot hold the total power together in a single pulse, thus it breaks into multiple pulses.

Surprisingly, a relatively newer TI material Tellurene has proclaimed the minimum saturation intensity of $\sim$1.06 MW/cm$^2$ \cite{liu2020ultrathin}. 
Tellurene is the 2D form of Tellurium thus, it comprises of only a single layer of crystal structure, resulting in a low DOS and less energy levels leading to minimum saturation intensity and fast saturation.
Nearly bandwidth-limited pulses was recorded in a Bi$_2$Te$_3$-based tapered-fiber coated TISA having TBP of 0.323 \cite{yan20152}.Therefore, by comparing all the reports in this domain, we have clearly highlighted the best possible TISA materials and their fabrication techniques.

Modelocking parameters such as repetition rate, pulse energy, 3-dB bandwidth and modelocked regimes are a few important ones which dictate the performance of the ultrafast fiber laser. Repetition rate of the laser determines the overall switching capability which is helpful in high-speed optical communication. The time-bandwidth product, i.e., the product of pulse duration in time domain and bandwidth of the same pulse in frequency domain, is an important parameter which determines how close the pulse is to the transform limit. The magnitude and the sign of the chirp also indicates the same, i.e, higher the chirp, longer the pulse gets stretched. The gain bandwidth of the laser ultimately limits the minimum duration of pulse width. The magnitude and sign of nonlinearity and dispersion play key roles to shape a pulse. For example, in longer fiber cavity, one can expect very high dispersion which stretched the generated laser pulses. On the other hand, if one incorporates nonlinearity in the fiber cavity by using small core diameter fiber, microfiber, PCF, highly nonlinear fiber (HNLF), etc., it compresses the pulse in time. The magnitude and sign of nonlinearity and dispersion is not evenly distributed inside the laser cavity, which leads to various schemes of pulse generation. 
Theoretically, by permutation and combination, there are as many as 81 types of soliton-like pulse generation schemes described. However, some of the combinations have been tried experimentally, among them some familiar modelocking schemes in fiber lasers are: fundamental solitons, dissipative solitons, dispersion-managed solitons, self-similar pulse evolution (similariton), nonlinearity managed solitons, split solitons, bunched solitons, etc.  \cite{ilday2004theory}.

\begin{figure}[h!]
\centering\includegraphics[width=14cm]{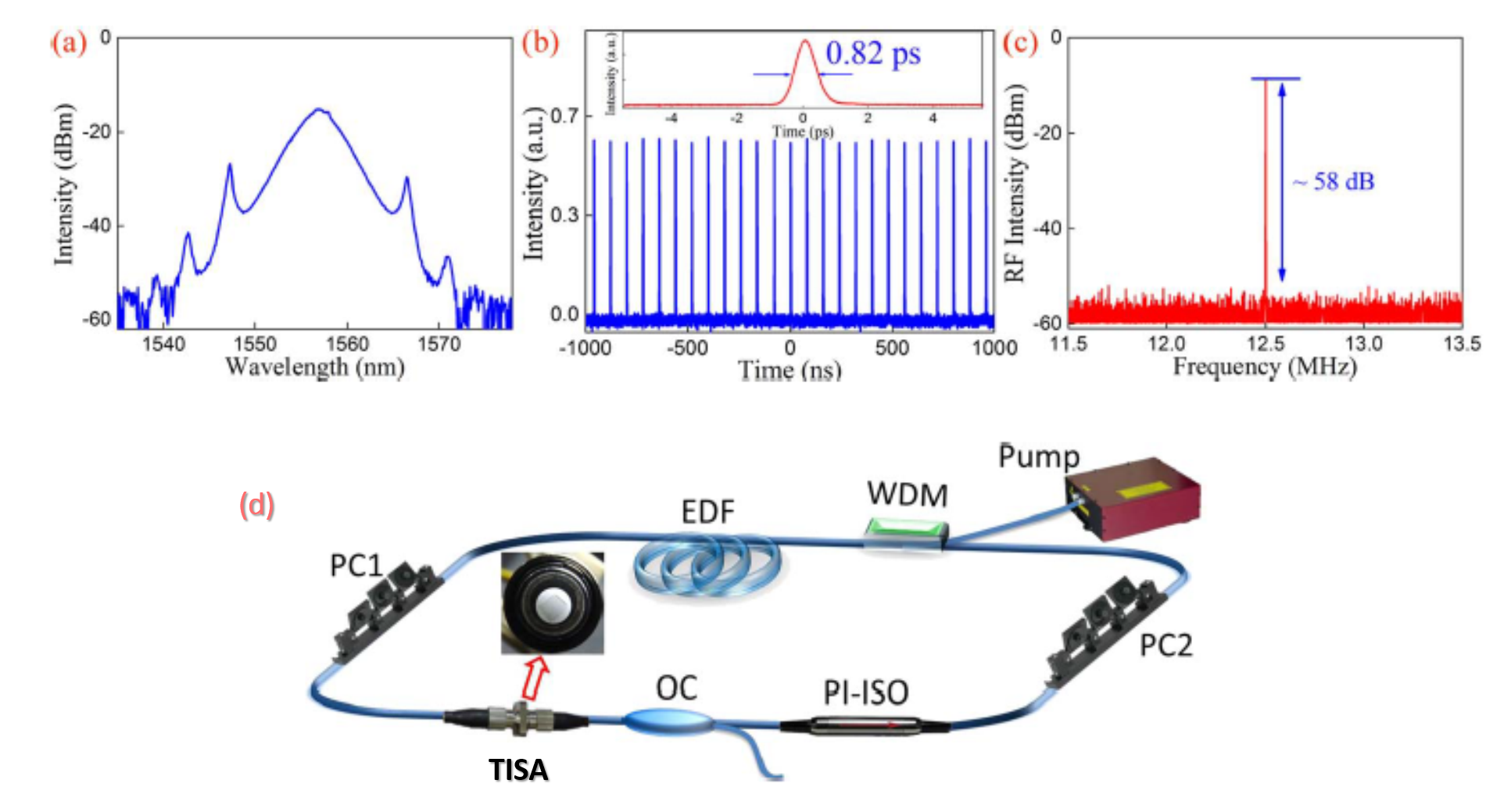}
\caption{Modelocked traces of fundamental soliton operation (a) Modelocked spectrum (b) Mode-locked pulse train;
inset, autocorrelation trace (c) RF spectrum and (d) Schematic of TISA-based passively mode-locked erbium-doped fiber laser. Reported with permission (Fig. 3,  \cite{luo2014observation}) 2014, IEEE}
\label{modelock_trace1} 
\end{figure}

\begin{figure}[h!]
\centering\includegraphics[width=14cm]{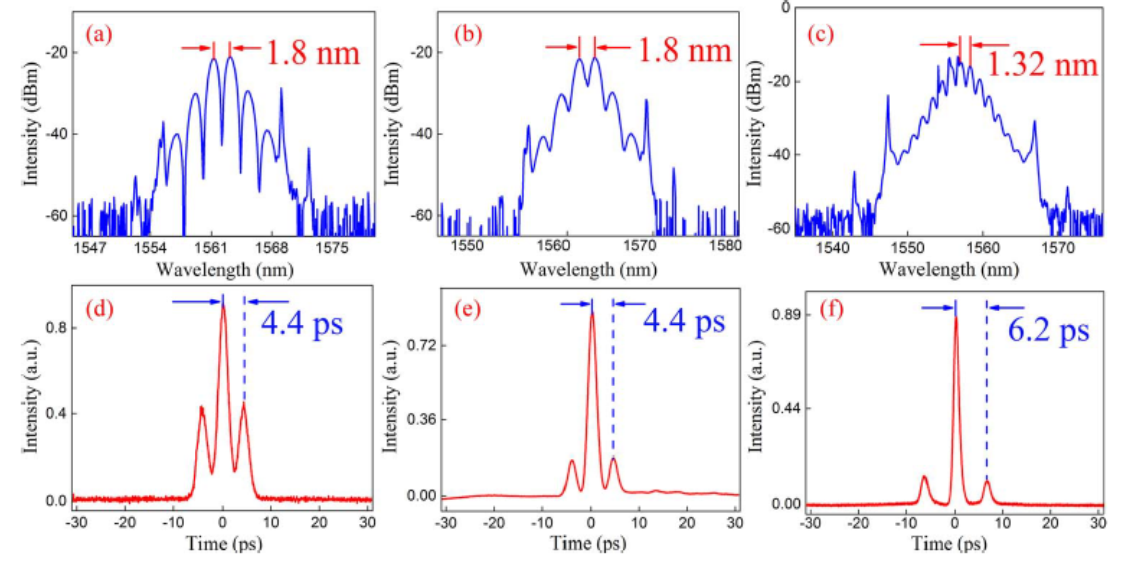}
\caption{Modelocked traces of three types of bound multi-pulse soliton states (a)–(c) Spectra; (d)–(f) Autocorrelation traces. Reported with permission (Fig. 4,  \cite{luo2014observation}) 2014, IEEE)} 
\label{bound-state-modelock}
\end{figure}

In fundamental soliton, the temporal and spectral shape of the pulse is conserved for infinite propagation distance. This happens when the self-phase modulation (Kerr nonlinearity) and chromatic dispersion inside resonator absolutely cancels each other. 
This fundamental soliton are achieved when moderate pump power has been used to the fiber laser cavity. Often it called as pure and stable modelocking regime which falls in the middle of Q-switched modelocked (QML) and pulse breaking boundary regimes. 
The fundamental modelocking traces such as modelocked spectrum, pulse train, autocorrelation trace, RF spectrum and the schematic diagram of TISA-based fiber laser cavity are shown in Fig. \ref{modelock_trace1} (a) to (d), respectively. 
Phenomena like chromatic dispersion and nonlinearity along with periodic disturbances are found in each resonator round-trip, which affects the soliton pulses. These disturbances couple the soliton with the dispersive wave which is primarily caused by the nonlinearity occurring from the discrete optical components. Generally, this coupling does not have significant effect in the fiber laser cavity due to the ever-changing values of the relative phase of soliton and the dispersive wave. However, in a fiber laser cavity, resonant coupling causes relative phase in multiples of 2$\pi$. This coupling generates narrow peaks in the optical spectrum which appear as Kelly sidebands. These Kelly sidebands are not visible in bulk lasers due to low nonlinearity and dispersion in the cavity. Thus, it can be concluded that Kelly sidebands (in Fig. \ref{modelock_trace1} (a)) are an indicator of mode-locked fiber laser cavity operating in soliton regime. 

When the pump power is lower than the QML threshold power, at the lower boundary regime (QML), the picosecond/femtosecond modelocked pulses are generated under the envelope of nanosecond Q-switched pulses. On the other side, when the pump power is relatively high and reaches at the boundary of pulse-breaking regime, multi-pulse solitons can be achieved. Depending on the level of intra-cavity peak power, fundamental soliton splits in two or multiple pulses and then unfold into the bound states. Fig. \ref{bound-state-modelock} shows three types of bound-state or multi-pulse soliton's spectrum ((a) -- (c)) and respective autocorrelation traces ((d) -- (f)). The inter-pulse distance should always correlate with the spectral modulation period if only bound states are present as seen from Fig. \ref{bound-state-modelock} (a) and (d). Also, higher the modulation depths, smaller will be the contrast between the peak intensities as long as the pulse separation is fixed at constant pump power (See Fig. \ref{bound-state-modelock} (a) \& (d) and Fig. \ref{bound-state-modelock} (b) \& (e)). However, when the pump power is too high, the inter-pulse distance also more and depth of modulation is also low (See Fig. \ref{bound-state-modelock} (c) \& (f)). 

Dissipative solitons (auto-solitons) refers to stable, nonlinear, time/space localized solitary pulses which result from the equilibrium between dissipation and energy excitation \cite{turitsyn2016dissipative}.
Similariton pulses generate from the fiber laser cavity when self-similar pulse generation occurs. Such kind of pulse generation arises when the pulse becomes independent of the shape and duration of the seed pulse, due to long propagation length and large amplification factor  \cite{oktem2010soliton}.
Dispersion managed solitons are generated in a fiber laser cavity 
exhibiting stretched pulses by employing normal as well as anomalous dispersion, through appropriate arrangement of fibers. Such solitons provide very high pulse energy compared to fundamental 
solitons \cite{turitsyn2012dispersion}.

\section{Challenges and Future Directions}
There are many challenges, but exciting results incurred with TISA based modelocked fiber lasers drive us to the future applications. The main challenge is to understand the electronic structure of TI material. Due to the strong spin-orbit coupling, these materials show complex electronic structures. Therefore, the theoretical prediction as well as the experimental measurement of the bandgap energies of these materials are difficult. Highly sophisticated measurement techniques are required to analyze the optical properties of TI. For the design of desired TISA based modelock fibre lasers, it needs expertise with state-of-art laboratory facilities. Moreover, the bandwidth of the pulse gets affected not only by the GVD and nonlinearity of the fibers, but also due to the presence of impurity, defect and dislocation in the material. Moreover, it is also observed that the bandgap energies of TISA varies with the number of quintuple layers. Hence, the synthesis and fabrication process should be optimized for the desired properties of TISA. When TISA fabricated with simpler techniques (such as, ME, LPE, HTE, etc.), the repeatability is a matter of concern. Due to imperfection, the non-saturable loss will be higher and undesirable.

From our investigation of TI-based mode-locked fiber lasers, several areas of research are found to be unexplored. So far, single TI based SA for different wavelength regimes is mainly reported in the literature. For the wavelength tuning and change of center wavelength, bandgap engineering with heterostructure of TI materials is a new direction in the ultrafast domain.  Moreover, TISA based ultrashort fiber lasers in the 2 $\mu$m regime are rarely reported in the literature. Moreover, among all fabrication techniques, PLD, which is one of the finest techniques for growing crystalline layered materials, have not been used to fabricate any kind of TISA at 2 $\mu$m wavelength. 

There are many mode-locking schemes, such as, dissipative solitons, dispersion-managed solitons, self-similar pulse evolution (similariton), nonlinearity managed solitons, split solitons, bunched solitons, all-fiber-all-normal-dispersion mode-locking, etc. are found in ytterbium and erbium fiber lasers, whereas, these schemes have not been studied properly in thulium/holmium-doped fiber lasers. 
Since thulium has a very wide gain bandwidth ($\sim$100 nm), one can expect much shorter duration than it has been reported ($\sim$230 fs \cite{chi2014all}) till now with TISA. Hence, extracting pulse having shorter duration at 2 $\mu$m wavelength is an open challenge to the researchers. It may be possible with a controlled fabrication technique in near future.

Moreover, from the literature survey it has been observed that MBE, grown TISA is not employed for mode-locking of fiber lasers at any wavelength. Note that, MBE is known to be one of the best techniques in terms of uniformity of monolayers providing control over the number of layers. However, MBE is a costlier method, used to grow thin film over a crystalline bulk substrate and mostly employed in a large-scale fabrication. Therefore, MBE-grown TISA is an open direction for investigation in fiber lasers.
\section{Conclusion}
In conclusion, a complete review of modelocked fiber laser using topological insulator based saturable absorbers is described. An in-depth and up-to-date literature review of all previously reported and concurrent works in this domain are depicted carefully with examples. 

The electronic and optical properties are obtained by quantum mechanical simulations for commonly used topological insulators and explained how these materials are useful as broadband saturable absorbers in fiber laser cavity at different wavelength regimes. The study also indicates that not only the 2D layers but also the few micrometer flakes can be used as an efficient saturable absorber. 
It is found that Bi$_2$Se$_3$ based TISAs are more appropriate for all available mid-IR fiber laser lines due to its unique band structures and available energy states. 

The synthesis and fabrication processes of TISA are described with diagrams. Among all fabrication techniques, pulsed laser deposition is found to be the most simple but efficient, uniform, reliable and controllable process of making TI films over various substrate and fiber positions. Amidst the all integration techniques, solution based TI nanoparticle filled photonic crystal fibers are found to be most efficient in terms of saturable absorption parameters. 

The techniques of surface, quality and optical characterization of TISA are reported. It is noticed that the oxidation degrades the quality of TISAs as it increases the relaxation time and decrease of the spectral-absorption bandwidth. However, the other parameters such as uniformity, impurity, inhomogeneity, etc. influences less to the saturable absorption properties. The open-aperture z-scan measurement is performed to derive the nonlinear optical property of Bi$_2$Se$_3$ TISA which matching to the previously reported works \cite{bhattachraya2016efficient}.

The basics of modelocking in fiber lasers using TISA are thoroughly investigated. The ability of TISA as a modelocker where dual or multi-wavelength modelocking is required, is discussed clearly. The comparison of all the previous works are tabulated with the necessary saturable absorption and modelocking parameters. The recipe of getting stable ultrashort modelocked pulses from TISA-based fiber laser is discussed. This study will definitely help to the application of ultrafast fiber lasers in mid-IR regime.       

\section{Acknowledgments} 
This work was supported by the 
SRM Innovation and Incubation Centre, SRM Institute of Science and Technology and funded by University Grant Commission, Govt. of India. KM would like to acknowledge the Department of Science and Technology for funding (DST/INSPIRE/04/2018/002482). The computational resources provided by ‘PARAM Shivay Facility’ under the National Supercomputing Mission, Government of India at the Indian Institute of Technology, Varanasi are acknowledged.
The author would like to acknowledge helpful discussions and fruitful collaboration with Dr. S. Chandramohan, Dr. E. Senthil Kumar, Dr. C Nayak (SRM IST, India), Dr. P K Datta (IIT Kharagpur, India) and Dr. S Ghosh (Georg-August-Universität Göttingen, Germany). The author acknowledges the Head, Department of
ECE, the Director of Engineering and Technology, and the Vice-Chancellor, SRM Institute of Science and Technology, Chennai, for their continuous encouragement.

\bibliographystyle{MSP}
\bibliography{sample}

\end{document}